\documentclass{aa}  

\usepackage{graphicx}
\usepackage{txfonts}
\begin{document}

   \title{Optimal parameter space for detecting stellar differential rotation and centre-to-limb convective variations}
    \titlerunning{Optimal parameter space for detecting DR and CLV}
    \authorrunning{Roguet-Kern et al.}

   \author{N. Roguet-Kern\inst{1}
          \and H.~M. Cegla\inst{2,3,1}\fnmsep\thanks{UKRI Future Leaders Fellow}
          \and V. Bourrier\inst{1}
          }
    \offprints{H. M. Cegla, \email{h.cegla@warwick.ac.uk}}

\institute{Observatoire Astronomique de l'Universit\'e de Gen\`eve, Chemin Pegasi 51b, CH-1290 Versoix, Switzerland
  \and Physics Department, University of Warwick, Coventry CV4 7AL, United Kingdom
  \and Centre for Exoplanets and Habitability, University of Warwick, Coventry CV4 7AL, UK
  }

   \date{Received <date> / Accepted <date>}

   \abstract{The reloaded Rossiter-McLaughlin method allows us to probe variations in the stellar surface by resolving spectra from the regions that are occulted by a planet as it transits. The goal of this paper is to investigate the optimal parameters space for using this technique to detect differential rotation (DR) and centre-to-limb convective variations. We simulated a star-planet system with and without convective effects to map the optimal regions of the parameter space for retrieving the injected differential rotation. Our simulations explored all possible ranges of projected obliquity (spin-orbit angle), stellar inclination, and impact parameter, as well as differences in instrumental configuration, stellar magnitude, and exposure time. We find that DR is more easily retrieved at low-impact parameters, corresponding to system configurations in which the transiting planet crosses the largest number of stellar latitudes. The main hot-spots for detection (i.e. areas in which DR detectability is high) are $120^{\rm{o}}<|\lambda|<180^{\rm{o}}$ for $i_*<90^{\rm{o}}$ and $|\lambda|<60^{\rm{o}}$ for $i_*>90^{\rm{o}}$ on average, and they tend to shrink as the impact parameter increases. Additionally, in contrast to the crucial impact of brightness, we identify that exposure time has a negligible impact on the difficulty of detecting DR as the increase in signal-to-noise ratio (S/N) at longer exposure times is counteracted by the degraded sampling rate. We determine that an ESPRESSO-like setup of instrumental configuration and sensitivity might retrieve DR up to $V=12$, compared to $V=10$ for HARPS.
   We reach no clear conclusion about limb-dependent convective effects and the possible confusion with DR; preliminary results suggest, however, that under certain circumstances, while it seems that one effect could be mistaken for the other, the accuracy of the fit (in particular of $\alpha$ ) does not hold up under additional scrutiny.}


   \maketitle
%

\section{Introduction}


    During a planetary transit, part of the stellar surface is eclipsed by the orbiting planet. This causes a distortion of the spectral lines from the disc-integrated star and a change in their associated radial velocities (RV) that is known as the Rossiter-McLaughlin (RM) effect. Because the RVs partly depend on the stellar rotation, the RM waveform is sensitive to the star-planet alignment, which provides information on the dynamical history of the system. For example, planets in aligned orbits may be indicative of a dynamically gentle planet-disc migration history, whereas misaligned planets may have experienced a mosre violent migration, such as planet-planet or star-planet scattering via the Kozai-Lidov mechanism \citep[and references therein]{triaud18}.

    If a star experiences differential rotation (DR) between the poles and the equator and/or convection in the stellar photosphere, these variations are naturally encoded in the line-of-sight velocities and are therefore also encoded in the RM signal. For example, our Sun rotates about 20\% faster at the equator than at the poles, which equates to a local change in velocity of $\sim$400~m~s$^{-1}$ across the full range of latitudes. Additionally, the convective envelope of our Sun means that the surface is corrugated: at disc centre, we see the very tops of the granules and the very bottoms of the intergranular lanes, but towards the limb, the tops and bottoms are obstructed from view and we begin to see the granular walls. Furthermore, some granules in the forefront of our line of sight obscure those in the background, and so on. The result are centre-to-limb variations (CLV) in a range of observed properties from continuum intensity to stellar line shape and net position or shift, with net velocity shifts of several hundreds of m~s$^{-1}$ \citep[see][and references therein, for more details]{lohner-bottcher18}. As spectrograph precision increases, ignoring these more subtle effects (as is often the case when modelling the RM effect) can inject systematic biases in our star-planet interpretations \citep{czesla15, cegla16a, casasayas-barris20, chen20}. Alternatively, if we account for these effects in RM modelling, we can overcome these biases and learn more about the physics on the stellar surface.   We therefore explore the optimal star-planet parameter space here in which transiting planets, in conjunction with the reloaded Rossiter-McLaughlin technique (RRM) \citep{cegla16b}, can be used to probe both stellar surface differential rotation and centre-to-limb (net) velocity variations induced by stellar surface convection.

    The RMM measures the spatially resolved stellar spectrum behind a planet as it transits, making it ideal for probing variations induced by the stellar surface. Using this approach, we can not only easily account for DR and limb-dependent convective effects, but also do not need to make any assumption about the local line shapes. This latter aspect is thanks to the direct subtraction of the in-transit data from out-of-transit observations (after the spectra have been normalised to a known transit light curve to compensate for the loss of absolute flux due to the Earth's atmosphere; see \citealt{cegla16b} for more details). Nonetheless, the orientation of the star-planet system also dictates which areas of the star are probed during the planet transit. Hence, some star-planet orientations are more or less favourable to detecting these subtle stellar surface effects.

    Naively, it is often assumed that systems in which the planet has a projected obliquity of 90$^{\rm{o}}$ are the most favourable for detecting stellar DR because the planet has a higher likelihood of transiting more stellar latitudes; there is some evidence to support this idea, as this was indeed the finding of \cite{serrano20}. However, \cite{serrano20} explored a relatively sparse parameter space (with the classical, disc-integrated RM modelling approach), simulating only four projected obliquities, three stellar inclinations, and five impact parameters. We here simulate a representative transiting hot Jupiter (similar to HD~189733~b) at all possible ranges of projected obliquity, stellar inclination, and impact parameter through a RMM analysis. Our goal is to determine the optimal parameter space for detecting and disentangling a solar-like stellar DR and CLV in the net convection-induced line-of-sight velocities.

    The foremost goal of our study is to determine in a first instance the star-planet system parameters that allow the easiest and most accurate detection of stellar DR. Additionally, we aim to determine the optimal parameter space in which limb-dependent variations in the net convective velocities may be isolated from the DR.

    In Section~\ref{Sect:sim} we present the model with which we simulate (and recover) the RRM signal from a transiting planet, including the introduction of stellar surface DR, limb-dependent net convective signal, and photon noise. Our attempt to probe the convection effects is explained further in Section \ref{Sect:CB}. We discuss our findings in Section~\ref{Sect:results} and conclude in Section~\ref{Sect:conc}.

\section{Simulation and model}
\label{Sect:sim}
    To accomplish our goal, we simulated a star-planet system during transit as observed from the Earth. In the optimal (empirical) cases, the target star is bright and the transit is well constrained by an existing light curve. For simplicity, we also assumed a circular orbit. We simulated a typical hot Jupiter because the high planet-to-star ratio of hot Jupiters boosts the signal-to-noise ratio (S/N) for the RRM analysis and their short orbital periods facilitate obtaining multiple transits. Additionally, they are also easier to observe within a single night, allowing for a generous out-of-transit baseline.

    We focused on an HD~189733-like system, as it is one of the brightest transiting hot Jupiter systems known to date and the first application of the RRM, wherein \cite{cegla16b} were able to loosely constrain the DR and CLV; $\alpha$ was found to be $> 0.1$ and $> 0.2$ with 99.2\% and 91.7\% confidence, respectively. Additionally, while the observations agreed with predictions from 3D MHD simulations, the observed local cross-correlation functions (CCF) did not exhibit significant convective variations. The following sky-projected obliquity and the stellar inclination were also recovered: $\lambda\approx-0.4\pm0.2^{\textrm{o}}$ and $i_{*}\approx92^{+12}_{-4}$ $^{\textrm{o}}$ (and a true 3D obliquity  $\psi\approx7^{+12}_{-4}$ $^{\textrm{o}}$).

    The three main parameters that we considered in our simulations are the stellar inclination $i_*$, the projected obliquity $\lambda$, and the orbital inclination $i_p$ (equivalent to the impact parameter $b$ for a fixed stellar radius and semi-major axis). Because of the way our model is set up, the planet always transits in front of the bottom half of the star from the perspective of an observer on the line of sight. Furthermore, we also studied the influence that the projected equatorial rotational velocity $v_{eq}\sin i_*$ and the visual magnitude $V$ have on our observations, as the RRM signal is amplified by a larger $v_{eq}\sin i_*$ and because the visual magnitude $V$ controls the S/N. Lastly, we briefly explore the impact of a difference in the cadence of observations on the results.

    As we follow the same coordinate system as \cite{cegla16b}, we varied $i_*$ from 0$^{\rm{o}}$ to 180$^{\rm{o}}$ and $\lambda$ from -180$^{\rm{o}}$ to 180$^{\rm{o}}$ in steps of 2$^{\rm{o}}$ and 4$^{\rm{o}}$, respectively. Between subsequent runs, we also varied the orbital inclination such that the impact parameter varied between 0 to 1 in steps of 0.05. Table \ref{tab:param} lists the system parameters that we held fixed for the HD~189733-like system in our simulations, including the limb-darkening coefficients $u_1$ and $u_2$.

    \begin{table}
        \centering
        \caption{Fixed parameters for HD 189733}
        \begin{tabular}{ccc}\hline\hline
            Parameters  & Value                     & Reference         \\\hline
            $R_*$       & 0.805~R$_{\odot}$         &  \cite{Boyajian15}\\
            $R_p$       & 0.15667~R$_{*}$           &  \cite{Sing11}    \\
            $a/R_*$     & 8.863                     &  \cite{Agol10}    \\
            $P$         & 2.21857567~d              &  \cite{Agol10}    \\
            $v_{eq}\sin i_*$ & 3.25~km~s$^{-1}$          &  \cite{cegla16a}  \\
            $V$         & 7.648                     &  \cite{Koen10}    \\
            $u_1$       & 0.816                     &  \cite{Sing11}    \\
            $u_2$       & 0                         &  \cite{Sing11}    \\\hline
        \end{tabular}
        \label{tab:param}
    \end{table}

    Our fitting loop operates as follows: For every orientation on a grid defined by $\lambda$, $i_*$ , and $b$ and using the fixed parameters described in Table \ref{tab:param}, the stellar velocity behind the planet at each point of the transit was calculated assuming a brightness-weighted solar DR law, following \cite{cegla16b}. These data served as our reference point, that is, they correspond to the true RVs behind the planet.

    To generate realistic velocity uncertainties for the local, occulted regions along the transit chord, we devised a model derived from the relation between surface RVs and the corresponding planet-occulted CCFs (independently of specific observations). The scaling of this formula is based on the local velocities determined from empirical observations of the HD~189733~b transits analysed in \citet{cegla16b}. The precision of the local RV is dictated by the amount of signal behind the planet, by instrumental sensitivity, the brightness of the star, and the exposure time. The formula is also scaled by a constant offset C, associated with instrumental setup and environmental conditions (that we assumed constant). Hence, our local RV uncertainties $e(RV,\mu)$ are defined as

   \begin{equation}
   \label{eqn:scale}
       e(RV,\mu) = \sqrt{\frac{1}{gain}\cdot\left(\frac{1}{R_*}\right)^2\cdot10^{\frac{V}{2.5}}\cdot\frac{1}{t_{exp}}}\cdot\frac{LC(\mu)}{1 - LC(\mu)}\cdot C,
   \end{equation}

    where the gain was set to 1 for HARPS and to 6 for ESPRESSO (equivalent to $flux'/flux,$ and it assumes the same exposure time), $R_*$ is the stellar radius, $t_{exp}$ is the exposure time in seconds, and the light curve (LC) is a function of $\mu=\cos\theta$, with $\theta$ the centre-to-limb angle.

    To simulate photon noise, we added Gaussian noise to the local velocities with a standard deviation equal to the value of the uncertainties, scaled from the empirical HARPS observations using Equation~\ref{eqn:scale}. For reference, the local velocity uncertainties in the HD~189733 HARPS observations were close to 50~m~s$^{-1}$ near disc centre and a few hundred m~s$^{-1}$ near the limb \citep{cegla16b}. To restrain the scope of the study, we did not consider any other sources of noise (instrumental or stellar) in our model. The impact of pressure-mode oscillations for Sun-like stars are likely to be largely mitigated with the optimal exposure time and/or binning \citep{chaplin19, hodzic20}; however, the effects of temporal convection evolution (i.e. granulation) is likely to increase the difficulty of recovering DR and CLV and will be examined in a forthcoming study. Throughout this analysis, we injected a solar-like DR, with a differential rotation rate $\alpha = 0.2$. An example of the simulated observations for a given orientation is shown in Figure~\ref{vstelnocb} by the black markers.

\subsection{Considering only differential rotation}
 \label{Sect:DR}

    In the first part of our study, we restricted our analysis to the sole contribution of DR (without CLV). The simulated data were fit using two different models: one model assumed DR, and the other assumed solid-body rotation (SB). In both models, we held the impact parameter $b$ fixed as we recognised that allowing this parameter to float freely may lead to degeneracy between fitting parameters. Throughout this paper, we place ourselves in idealised cases to better understand the detectability of CLV and DR with system properties (i.e. we assume the brightest systems, with the best light curves available). The DR model has four free parameters: the projected obliquity $\lambda$, the stellar inclination $i_{*}$, the equatorial velocity $v_{eq}$, and the differential rotation coefficient $\alpha$. When we fit the SB model, we had two free parameters: $\lambda$ and $v_{eq}{\sin i_{*}}$. In this case, the parameters $i_{*}$ and $\alpha$ were fixed to 90$^\textrm{o}$ and 0, respectively (such that $v_{eq} = v_{eq}{\sin i_{*}}$). In both models, we used non-linear least squares to fit the data. The fitting algorithm needs a set of starting parameter values, one for each free parameter, which we provided as a random value within a certain range centred on the true value (typically half the possible range for a given parameter). The fitting process allows each free parameter to take any value within the bounds of the parameter space, independently of the value of any other free parameter. Sections~\ref{subsubsec:accuracy_DR} and \ref{subsubsec:accuracy_CLV} provide more detail about the accuracy of the fits.

\begin{figure}
    \centering
    \includegraphics[width=8.5cm]{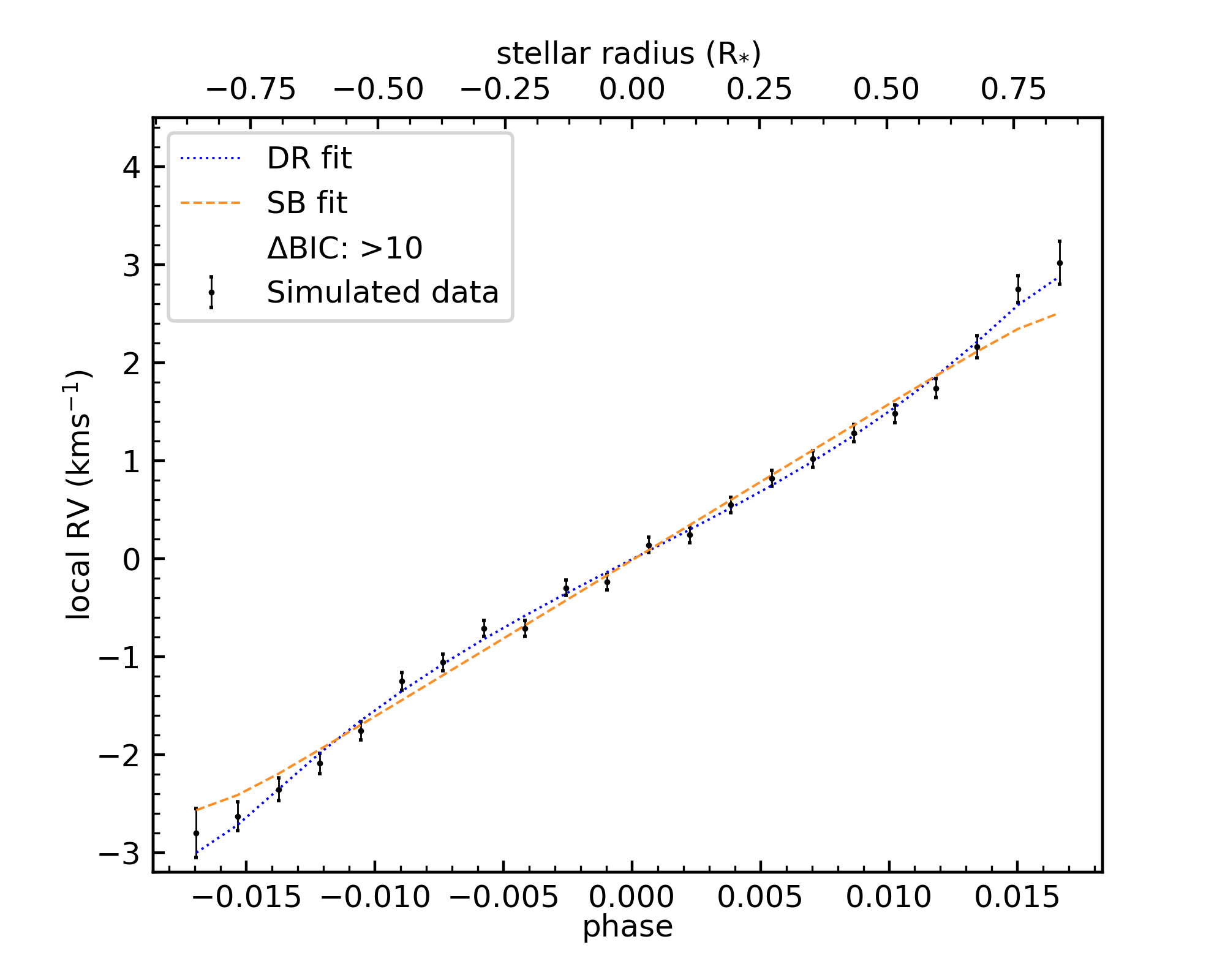}
        \caption{Simulated local stellar surface velocity below the planet along the transit chord as a function of the phase (bottom) and stellar radius (top) for a given star-planet orientation ($\lambda = 0^{\rm{o}}; i_* = 150^{\rm{o}}; b=0.2; V=10$). The best-fit DR model is shown in blue, and the best-fit SB model is plotted in orange; a positive BIC indicates a preference for the more complex DR model. }
        \label{vstelnocb}
    \end{figure}

    To compare the DR and SB fits, we first calculated the $\chi^2$ for each. Computing the Bayesian information criterion (BIC) for each fit allowed us to prefer the model with the lower of the two values whilst being conscious of overfitting the data with too many free parameters. A difference $\Delta BIC$ $>2$ is already considered positive, while $\Delta BIC$ $>6$ indicates a strong preference for the smaller BIC model \citep{Kass95}. Hence, in the graphs presented in this paper, any difference in the BIC value higher than 10 is rounded down to that number to facilitate interpretation. Figure \ref{vstelnocb} shows an example of an RV curve, with the simulated data and the two fits using our DR and SB models.

\subsection{Considering a centre-to-limb variation in convection}
 \label{Sect:CB}
    The second goal of our study is to investigate the impact of CLV on the detection of DR and to determine when these two effects may be differentiated; we use CLV hereafter to refer to the net limb-dependent convection-induced radial velocity shifts. To do this, we injected limb-dependent convective velocities into our previously simulated observations and tried to fit a variation of our two models on this adjusted input.

    The injected convective velocities were derived from the magnetically quiet (net 0~G) solar simulations presented in \cite{cegla18}. Therein, the authors constructed 3D magnetohydrodynamic (MHD) simulations and synthesised the Fe I 6302 \r{A} absorption line profile. The MHD simulations were inclined at nine centre-to-limb positions in steps of 0.1~$\mu$ and corresponded to six snapshots separated in time by 20 minutes each. To emulate exoplanet observations that use CCFs, the limb-dependent velocities were determined by cross-correlating all line profiles to a single profile at disc centre and taking the centroid of these resulting CCFs. Hence, these velocity shifts encapsulate all centre-to-limb variations resulting from the 3D stellar atmosphere; this includes both net shifts from which velocity flows are visible along the line of sight and shifts arising from changes in the line profile shape or asymmetry. As constructed, the velocities are relative to disc centre and thus do not encode the net convective blueshift present at disc centre; this is by design as the disc-integrated net convective blueshift is subtracted off by construction in the RRM (when the systemic and nightly radial velocity offsets are removed). We averaged these velocities over the time series to isolate the net changes across the limb, and fit a fourth-order polynomial to extrapolate to any limb position. The convective velocities were then simply added to the simulated stellar rotational velocities from Section \ref{Sect:sim}, denoted hereafter as either DR+CLV or SB+CLV.

    Since the MHD simulations are based on a solar twin and only one line profile was synthesised (albeit a typical line, representative of many lines that dominate most CCF template masks), the injected convection model might not be entirely representative of the velocities from a large spectral bandpass or for other stellar types. Moreover, because we used a polynomial fit to a time-average of snapshots, we did not account for temporal variations here. Nonetheless, this investigation provides a starting ground for estimating our ability to recover a limb-dependant signal.

    To recover the CLV in the simulated observations, we considered two more free parameters (in addition to those described in Section~\ref{Sect:sim}): the coefficients $c_1$ and $c_2$ of a second-order polynomial that depends on the centre-to-limb angle; the polynomial coefficient $c_0$ was derived mathematically from $c_1$ and $c_2$ (see \citealt{cegla16b} for details). We dropped the third and fourth coefficients in the fitting procedure as we found that with the level of injected photon (Gaussian) noise, we would be overfitting the data (this is consistent with empirically reloaded RM analysis in the literature); this simplification also increased the computational speed.

    \begin{figure}
    \centering
    \includegraphics[width=8.5cm]{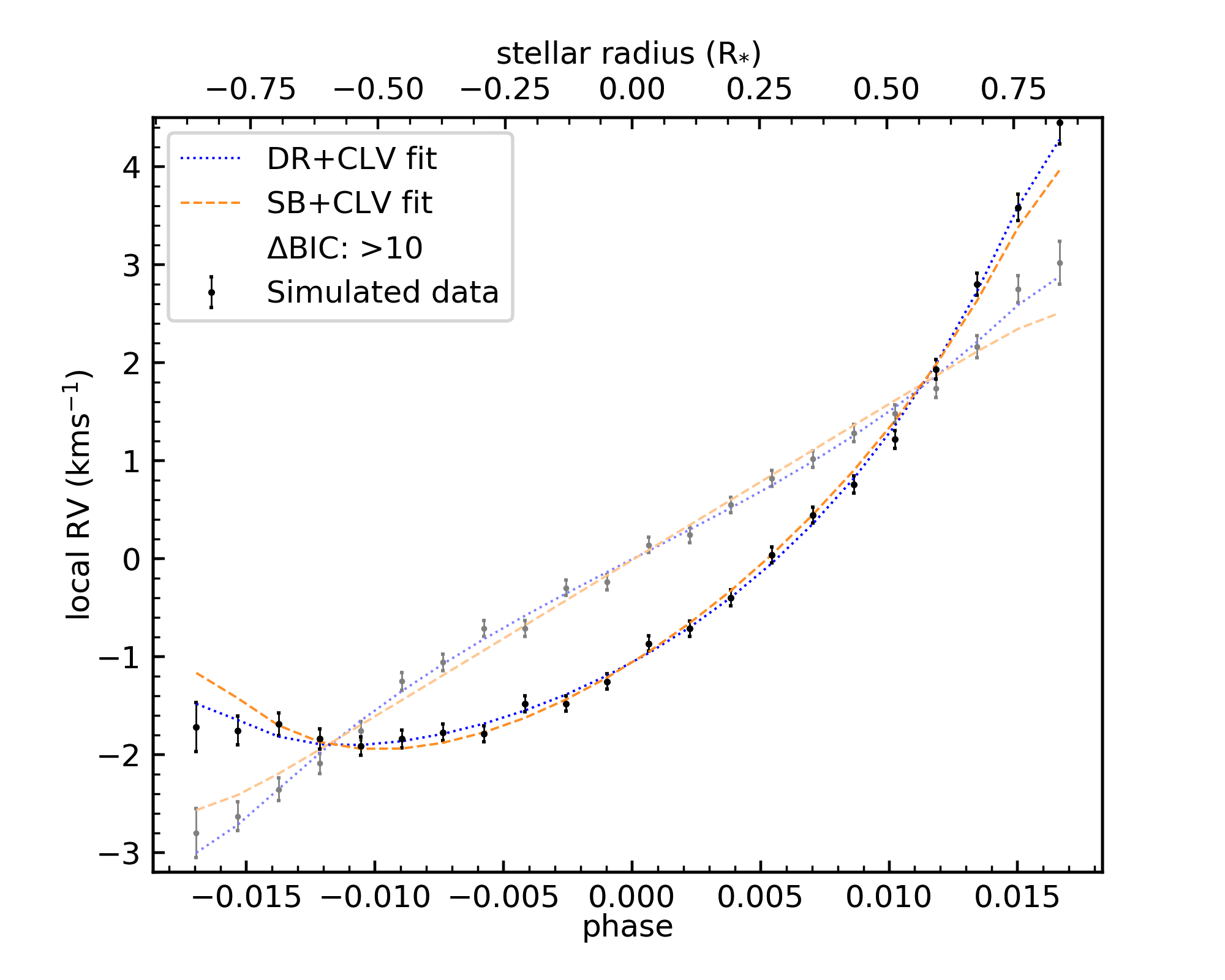}
        \caption{Simulated local stellar surface velocity below the planet along the transit chord as a function of the phase (bottom) and stellar radius (top) for a given star-planet orientation ($\lambda = 0^{\rm{o}}; i_* = 150^{\rm{o}}; b=0.2; V=10$), wherein both a solar DR and CLV are injected; compare to Figure~\ref{vstelnocb} to note the CLV impact (overplotted here in light colours), particularly the relative redshifts near the limbs. The corresponding DR+CLV model fit is shown in blue, and the CLV+SB fit is plotted in orange; a positive BIC indicates a preference for the more complex DR+CLV model.
        }
        \label{vstelcb}
    \end{figure}

    In Figure \ref{vstelcb} we show an example of one star-planet orientation where we included the convective contribution in the reference data. All other parameters being equal, the shape of the curve is clearly radically different from the plot in Figure \ref{vstelnocb}, where no convective effects are considered. The largest differences can be noted at the limbs, where the convection contributes towards a relative redshift; although the net convective effect at disc centre is a blueshift of a few hundred m~s$^{-1}$, which dramatically decreases towards the limb, this is a relative effect that is largely subtracted off when we remove the systemic velocities in the empirical observations.

    It is important to note that fitting with CLV increases the number of free parameters, which is heavily penalised in the BIC value. We recall that a positive change in the BIC indicates a preference for the more complex model, and because a $\Delta BIC$$>2$ is already significant, we do not give precise values above $\Delta BIC$$=10$ to facilitate visual interpretation of the graphs.


\section{Results and discussion}
\label{Sect:results}
    In the following section, we present our findings for the optimal parameter space for detecting and disentangling DR and CLV using the RRM technique. First we present our results when stellar rotation alone was considered. Then we present our analysis when a solar-like DR and CLV are considered. To help determine the optimal parameter space for which a DR fit is preferred to SB (or DR+CLV versus SB+CLV), we display our results in two and three dimensions, using colour-coded heat maps.

    The 2D heat maps are retrieved by holding the impact parameter $b$ constant while we varied the stellar inclination $i_*$ (on the vertical axis) and the stellar projected obliquity $\lambda$ (on the horizontal axis). They represent the model preference over all star-planet orientations considered for a particular value of $b$ and are colour-coded based on the BIC difference between the DR and the SB fit models; a darker shade of green (i.e. more positive $\Delta BIC$) indicates a significant preference of the DR fit (or DR+CLV) over the SB fit (or SB+CLV). On the other hand, light green and white indicate that DR (or DR+CLV) cannot be preferred over SB (or SB+CLV) with confidence. All negative $\Delta BIC$ returns in which SB is preferred were set to zero in the plots to more easily identify the regions in which a DR detection is possible. As a reminder, $\Delta BIC$ values higher than 10 were set to 10 to facilitate the presentation of the results in the figures.

    Comparison between the various 2D heat maps gives a clear representation of the potential of retrieving DR (or DR+CLV) as the impact parameter increases. To provide a convenient way to compare the results for multiple values of $b$, we generated a 3D representation by stacking the 2D plots over the whole range of impact parameters. To facilitate the interpretation, we projected the 3D structure that emerges onto the corresponding planes. The label "fill" associated with the 3D plots provides a way to quickly appreciate the fraction of the parameter space below the top surface that holds orientations favourable to the detection of DR. Technically, because the parameter space is divided into cells for which we know the preference for DR over SB, it is trivial to average this value over a whole stack of equivalent orientations (varying only the impact parameter). A lighter colour therefore corresponds to the presence of holes in the stack below the top surface,  which is defined as the highest impact parameter for which the DR fit is preferred over the SB fit. The colour scale of the projected contours on the two vertical planes corresponds to the fraction of the orthogonal stack in the parameter space for which DR is preferred.

\subsection{Stellar rotation alone}
\label{subsec:Rot_Only}

    We compared DR and SB alone (i.e. without CLV). We recall that the DR model has four free parameters ($\lambda$, $i_*$, $v_{eq}$ , and $\alpha$) and the SB model has only two ($\lambda$ and $v_{eq}\sin i_*$). Initially, we set the injected $v_{eq}\sin i_*$ and $V$ magnitude according to the values listed in Table~\ref{tab:param}. We plot a series of 2D heat maps at increasing values of $b$ to probe the DR preference in the parameter space.

    \begin{figure*}[h]
    \centering
    \includegraphics[trim=0mm 0mm 0mm 11.1mm,clip,width=6cm]{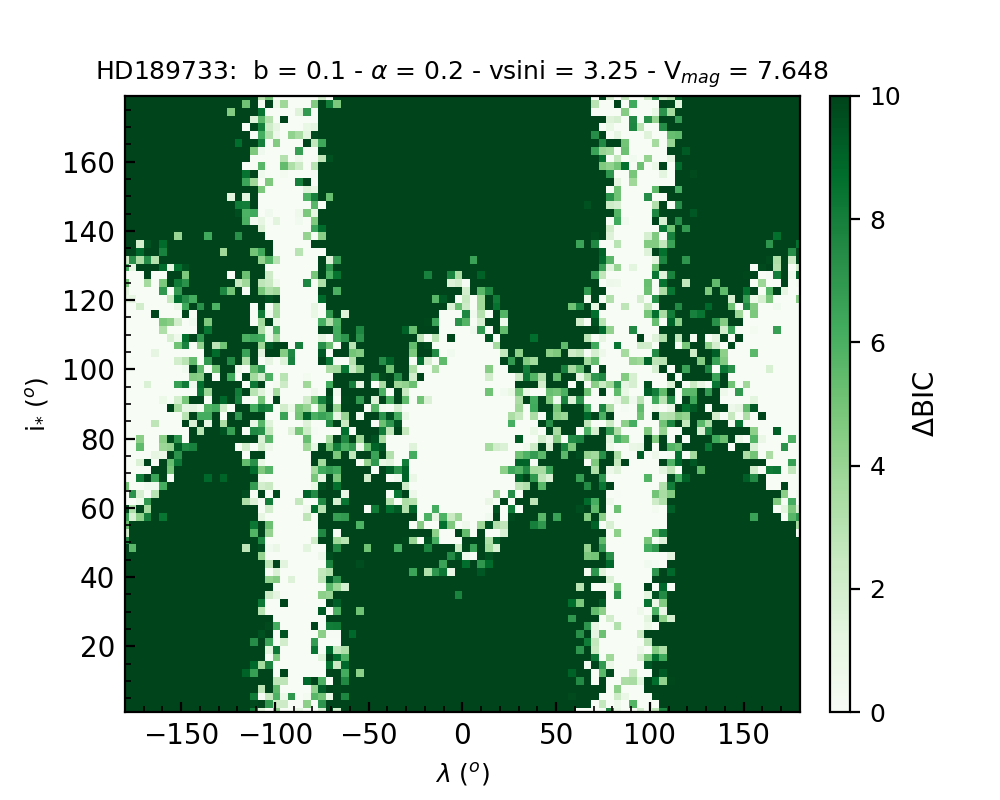}
    \includegraphics[trim=0mm 0mm 0mm 11.1mm,clip,width=6cm]{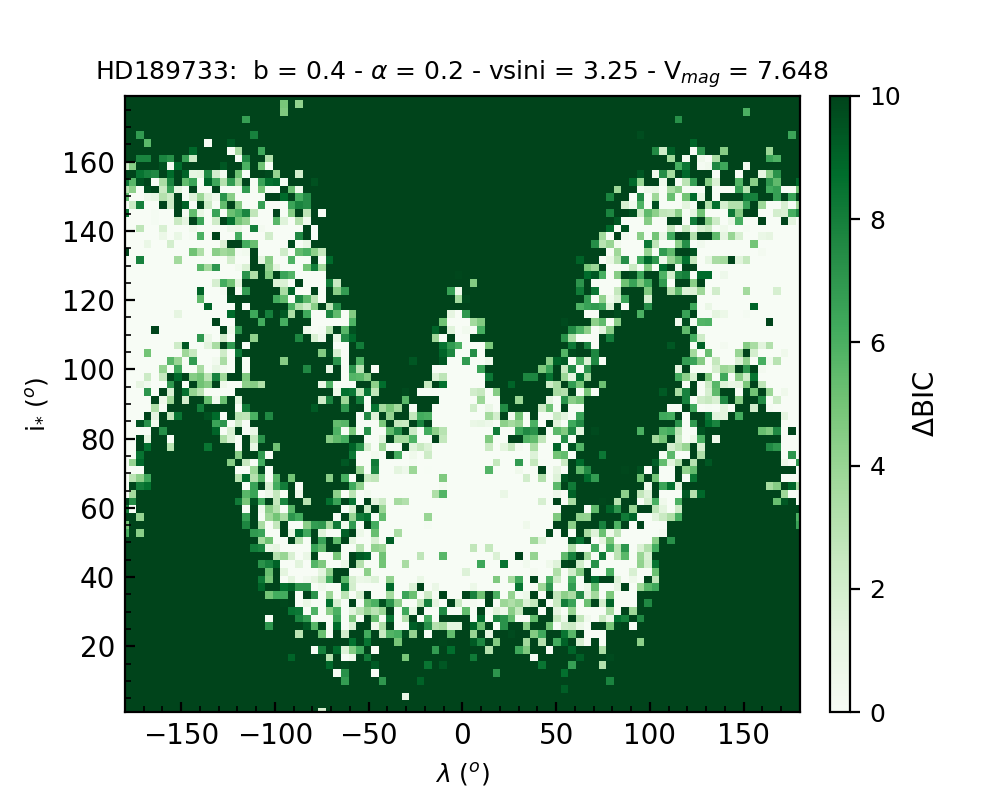}
    \includegraphics[trim=0mm 0mm 0mm 11.1mm,clip,width=6cm]{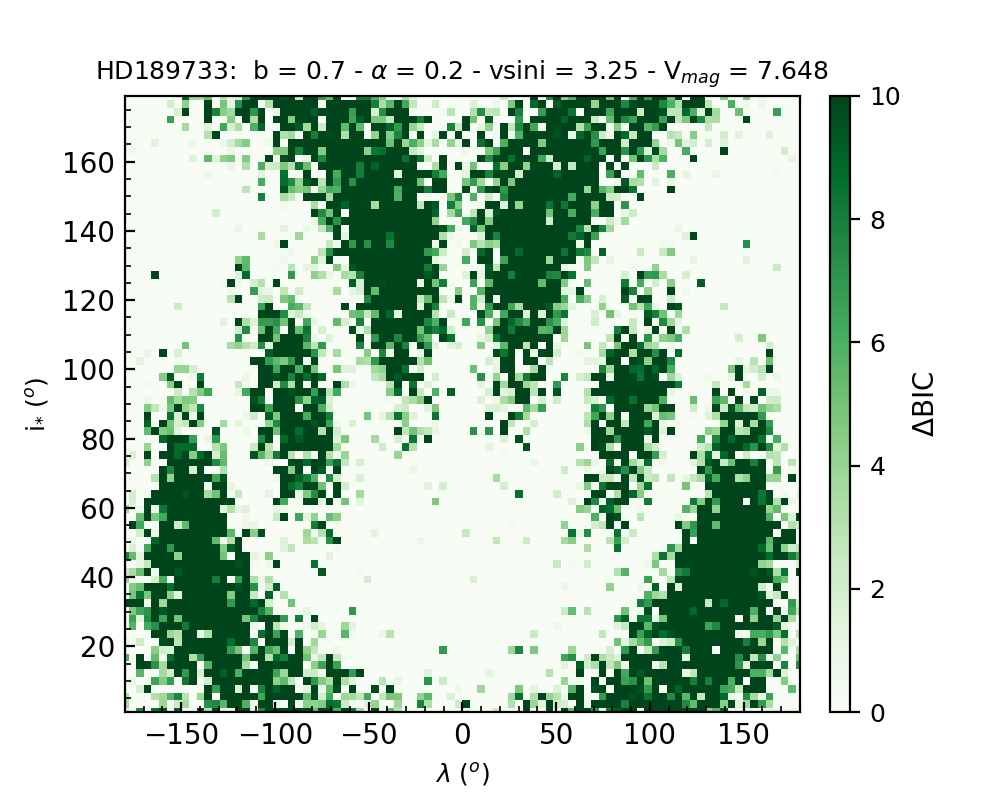}
        \caption{
        Two-dimensional heat maps illustrating the preferred parameter space (colour-coded by the change in BIC) for detecting stellar DR as a function of projected obliquity and stellar inclination for three different impact parameters ($b = 0.1, 0.4,\text{ and } 0.7$ from left to right). For a given system, as the impact parameter increases, the parameter space for preferring a DR model over SB decreases and isolated hot-spots appear. These results have been obtained with $\alpha=0.2$, and the parameters that were held constant in the three figures are listed in Table~\ref{tab:param}.
        }
        \label{hd189nocbnoalpha}
    \end{figure*}

    As Figure~\ref{hd189nocbnoalpha} demonstrates, for low values of $b$, most of the parameter space allows a clear detection of DR, except in a few areas in which $i_*=90^{\rm{o}}$. Most of these regions of poor DR detection correspond to stellar orientations in which the transiting planet crosses the fewest stellar latitudes. We also note two white streaks in the left side plot at $\lambda\approx\pm90^{\rm{o}}$, which is a sign of a small degeneracy. At $\lambda = \pm90^{\rm{o}}$, the planet occults only one longitude value, and at b=0, this means that the planet crosses a region of zero stellar rotation; hence, at low impact parameters and perpendicular projected alignments, the occulted stellar rotation can fall below the measurement precision. It is also important to note that for low-impact parameters, there can be degeneracies in the observed transit light-curve fit, such as between $a/R_{\star}$, $i_p$ and $T0$; hence, a more precise light curve may be needed to retrieve the results empirically. Using ESPRESSO and its higher precision, we note that the streaks of Figure~\ref{hd189nocbnoalpha} narrow down and even fill in around $i_*\approx90^{\rm{o}}$ (Figure~\ref{hd189733fullvmag10H} and Figure~\ref{hd189733fullvmag10E} are useful to visualize the difference in sensitivity between a HARPS and an ESPRESSO setup, with the latter drastically improving detection compared to the former).

    The picture changes, however, as the impact parameter increases, and our significant preference for DR decreases rapidly. At high values of $b$, DR can only be reliably detected in distinct areas of the parameter space, listed below:

    \begin{itemize}
        \item $i_*<90^{\rm{o}}$ and $120^{\rm{o}}<|\lambda|<180^{\rm{o}}$
        \item $i_*>90^{\rm{o}}$ and $|\lambda|<60^{\rm{o}}$
        \item $i_*\approx90^{\rm{o}}$ and $\lambda\approx\pm90^{\rm{o}}$.
    \end{itemize}

    Around these hot-spots (i.e. areas where DR detectability is high), the DR model is preferred over SB up to high values of the impact parameter $b$. The hot-spots likely arise because they correspond to configurations in the system in which the planet crosses a maximum of latitudes on the star, that is, regions in which the effects of the differential rotation are heightened. When we map the range of latitudes that the planet crosses during the transit, we indeed observe a maximum around $\lambda=\pm90^{\rm{o}}$ and $i_{*}=90^{\rm{o}}$ , regardless of the impact parameter. Furthermore, as the planet always transits in front of the lower half of the star in our model (due to our coordinate system), regardless of the stellar orientation, the range of latitudes crossed is higher when the star is farther away from the pole ($i_*$ > 90$^{\rm{o}}$) than when it is pole-on ($i_*$ < 90$^{\rm{o}}$), which explains the vertical asymmetry in Figure~\ref{hd189nocbnoalpha}. This effect is more prevalent at lower impact parameters. At higher impact parameters, the majority of the transits occurs near the limbs of the star, where the S/N is worse. Another major factor is the (assumed) instrumental precision, as an ESPRESSO/VLT-like setup is much better at picking up DR at high values of $b$ than a setup with HARPS-like precision (cf. Figure~\ref{hd189733fullvmag10H} and \ref{hd189733fullvmag10E}). Other factors that affect our ability to retrieve DR at a high impact parameter are the absolute magnitude of the rotational shear $\alpha$, the projected stellar rotational velocity $v_{eq}\sin i_*$, and the sampling rate of the observations, for example, as we discuss in Sections~\ref{subsubsec:sampling_rate_nocb} and \ref{subsubsec:rotation_rate_nocb}.

\begin{figure}[h]
    \centering
    \includegraphics[trim=3.5cm 1cm 1.4cm 1.5cm, clip, scale=0.43]{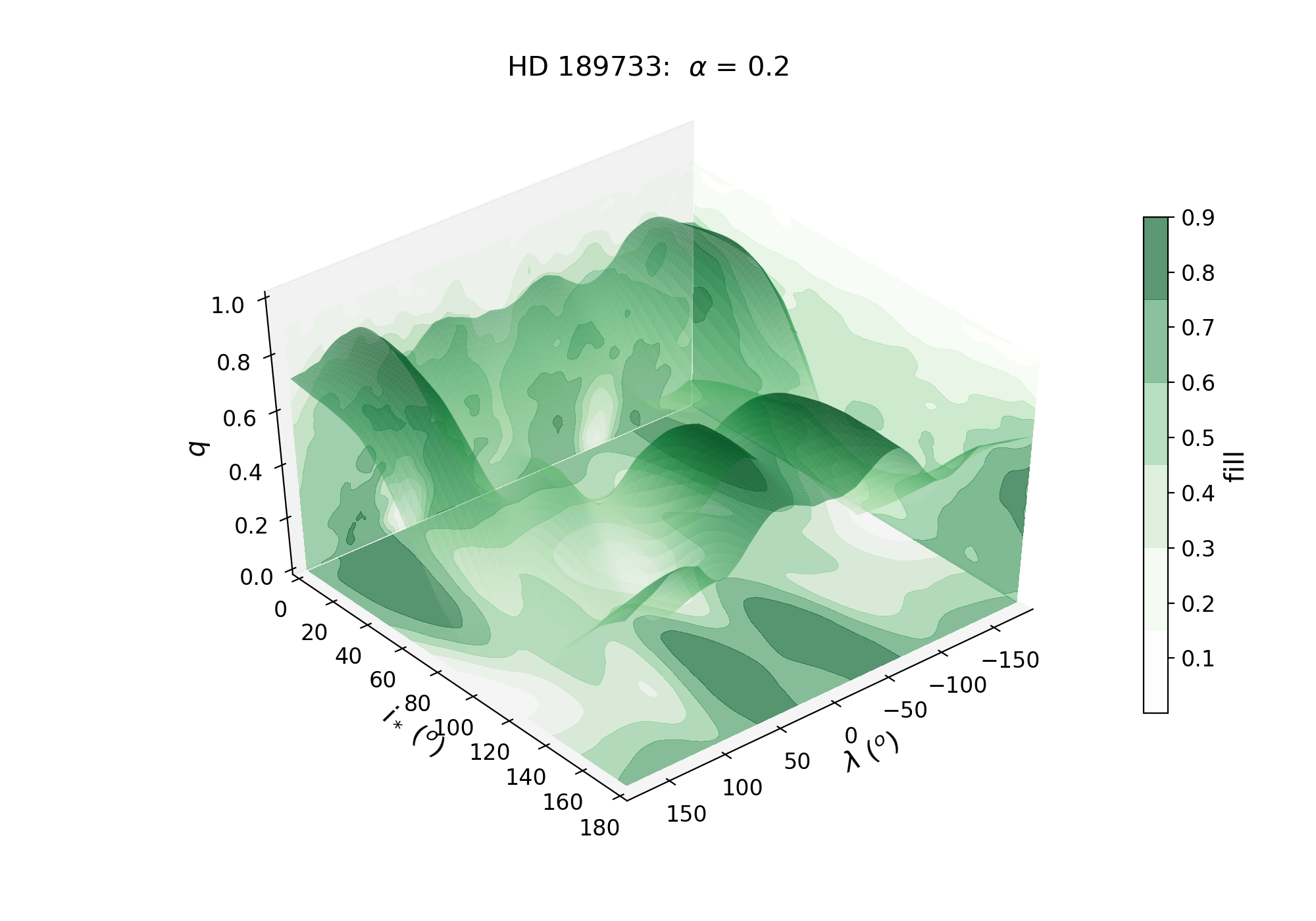}
        \caption{
        Three-dimensional heat map of the likelihood of detecting DR, $\alpha=0.2$. The parameters held constant are listed in Table~\ref{tab:param}. The 3D surface represents the highest impact parameter at which DR can be retrieved for the corresponding stellar orientation. The structure of the volume below this surface is represented by way of the projections on the respective planes and according to the colour bar. A darker colour indicates a lower preference for DR in the projected area.
        }
        \label{hd189733fullv3.5}
    \end{figure}

   To give a point of comparison in the subsequent discussion, we present the 3D parameter space for a HD~189733-like system in Figure~\ref{hd189733fullv3.5}. This figure corresponds to the stacking of several 2D plots like those presented in Figure~\ref{hd189nocbnoalpha} and therefore presents the same features.

    Figure~\ref{hd189nocbnoalpha} and  Figure~\ref{hd189733fullv3.5}  demonstrate that, in theory, a large parameter space is expected in which we may prefer a DR model fit over a SB scenario. However, these figures do not convey the accuracy of the DR fit as there are no constraints on the rate of DR, that is, $\alpha$. Hence, it is possible that an incorrect DR model could be preferred and therefore inject systematic biases into our understanding of solar-type stars.

    \subsubsection{Accuracy of fits}
    \label{subsubsec:accuracy_DR}

\begin{figure*}[h]
    \centering
    \includegraphics[trim=0mm 0mm 0mm 11.1mm,clip,width=6cm]{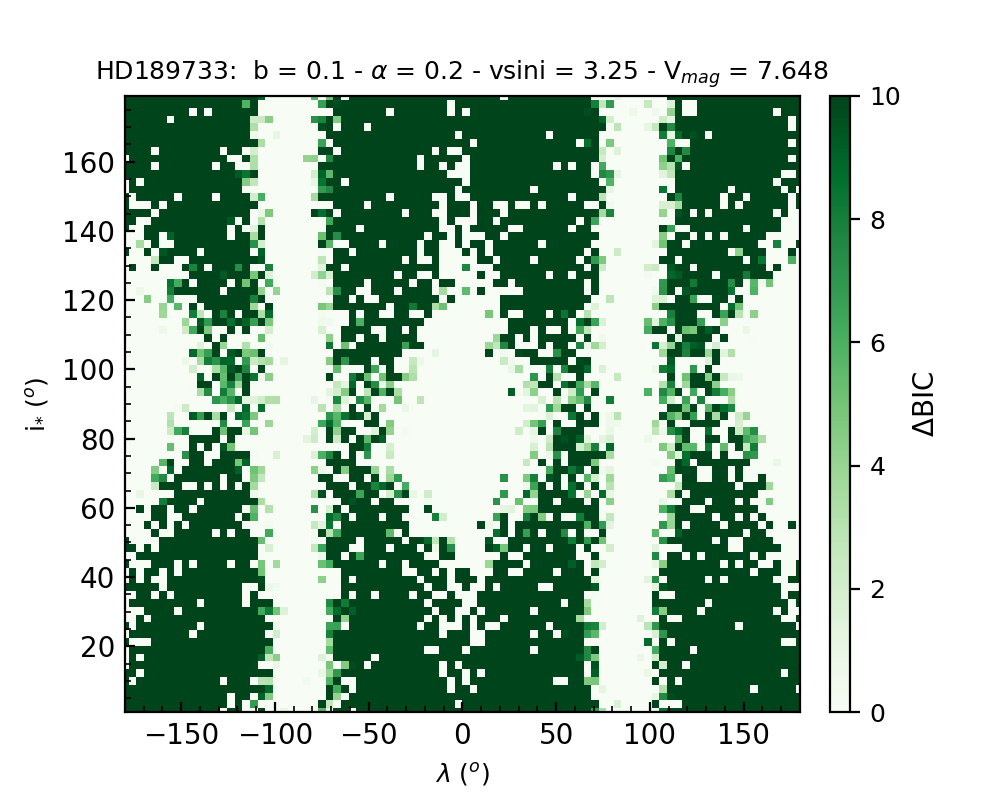}
    \includegraphics[trim=0mm 0mm 0mm 11.1mm,clip,width=6cm]{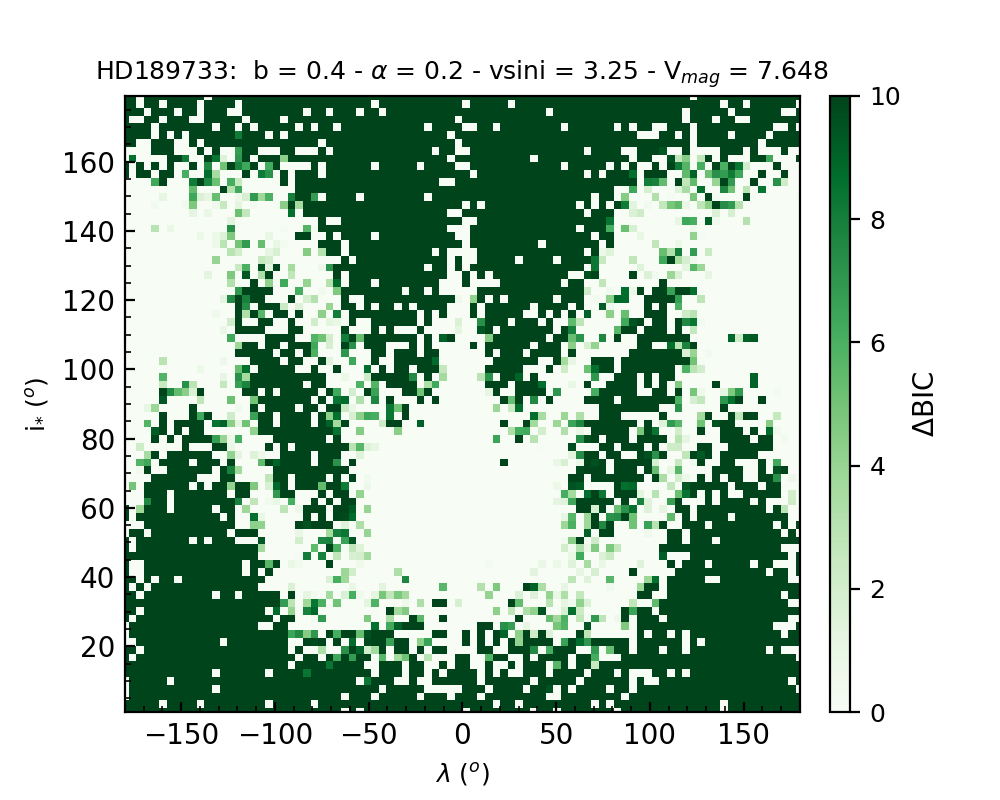}
    \includegraphics[trim=0mm 0mm 0mm 11.1mm,clip,width=6cm]{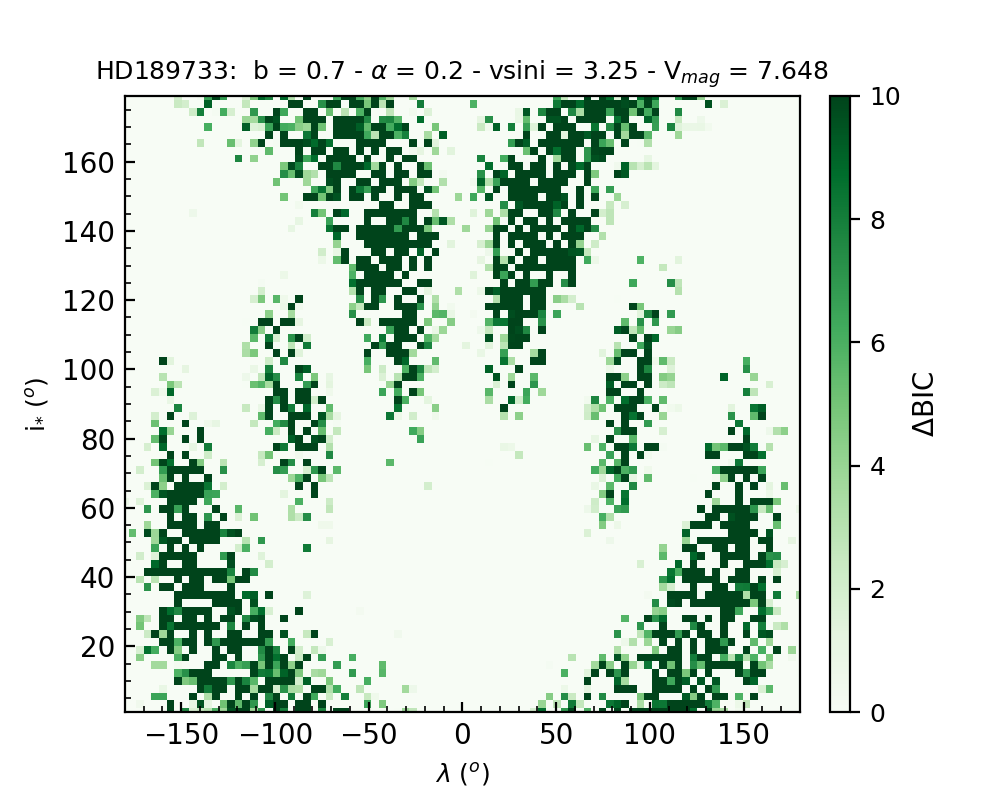}
        \caption{
        Two-dimensional heat maps illustrating the preferred parameter space (colour-coded by the change in BIC) for accurately detecting stellar DR as a function of projected obliquity and stellar inclination for three different impact parameters ($b = 0.1, 0.4,\text{and } 0.7$ from left to right). Accurate detections are those for which the recovered $\alpha$ agrees with the injected value within 0.05. The behaviour is similar to Figure~\ref{hd189nocbnoalpha}, where accuracy is ignored, but the cold spots are even larger; regions that change colour between these two figures should be taken with caution in empirical analyses. These results have been obtained with $\alpha=0.2$, and the parameters that we held constant in the three figures are listed in Table~\ref{tab:param}.}
        \label{hd189nocbalpha}
    \end{figure*}

    To investigate the accuracy of our DR fits, we highlight in Figure~\ref{hd189nocbalpha} only the regions in which both a DR fit is preferred and $\alpha$ agrees within 0.05 of the injected value. We find that the outcome is very similar, that is, the aforementioned distribution of hot spots is clearly visible. In particular, we observe that at low impact parameters, our ability to accurately constrain DR is only slightly worse (about $\lambda=0^{\rm{o}}$ and $\lambda=\pm 90^{\rm{o}}$), but it decreases significantly as the value of $b$ increases. Comparing the right plots in Figure~\ref{hd189nocbnoalpha} and Figure~\ref{hd189nocbalpha}, we observe a clear deterioration in the sharpness of the boundaries between high- and low-confidence areas (green and white areas). The areas in which DR was easily preferred previously are now more sparsely populated, which indicates that the DR fits are less reliable in these regions. Regarding the accuracy of the fits on $\lambda$, $i_*$, $v_{eq}$ and $\alpha$ (when such parameters are fitted), our results are remarkably close to the injected value for the most part. Starting with $\lambda$, the mean offset between the fitted value and the real value is only $0.10^{\rm{o}}$, with a standard deviation of $4.81^{\rm{o}}$: the small offset means that the fitted results are distributed equally around the injected value, with minimum bias. However, the width of the distribution means that the fit is not perfectly accurate. These values are slightly higher on $i_*$, with a mean offset between fitted and real values of $-0.20^{\rm{o}}$ and a standard deviation of $21.85^{\rm{o}}$. While the fitting accuracy on $v_{eq}$ looks worse with a mean offset of $-6.71$ km~s$^{-1}$ and a standard deviation of $39.41$ km~s$^{-1}$, when we reconstruct the value of $v_{eq}{\sin i_{*}}$ , we find that the predicted value has a mean of $3.36$ km~s$^{-1}$ (standard deviation: $1.76$ km~s$^{-1}$) that is surprisingly close to the injected value of $3.25$ km~s$^{-1}$ (cf. Table~\ref{tab:param}). Regarding $\alpha$, we find a mean offset of $-0.09$ and a standard deviation of $0.26$. As Figure~\ref{hd189nocbalpha} illustrates, the fitted values on $\alpha$ show a higher dispersion than the injected parameter, especially at a higher impact parameter.

    \subsubsection{Impact of the exposure time}
    \label{subsubsec:sampling_rate_nocb}

\begin{figure*}[h]
    \centering
    \includegraphics[trim=0mm 0mm 0mm 11.1mm,clip,width=8.5cm]{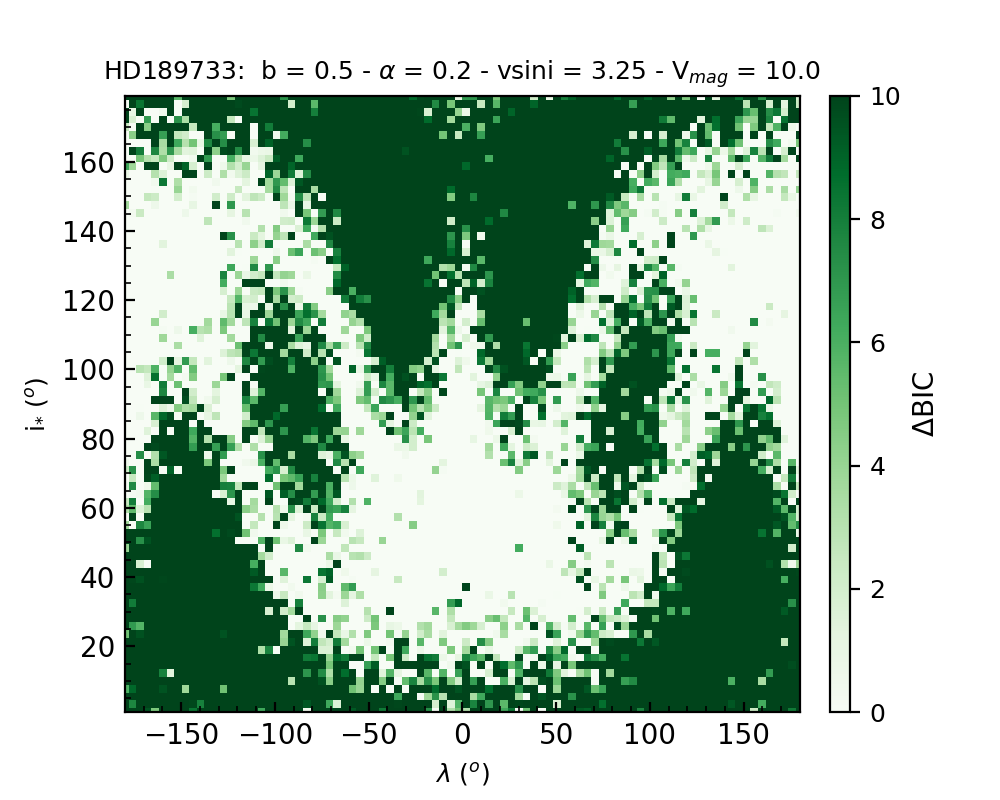}
    \includegraphics[trim=0mm 0mm 0mm 11.1mm,clip,width=8.5cm]{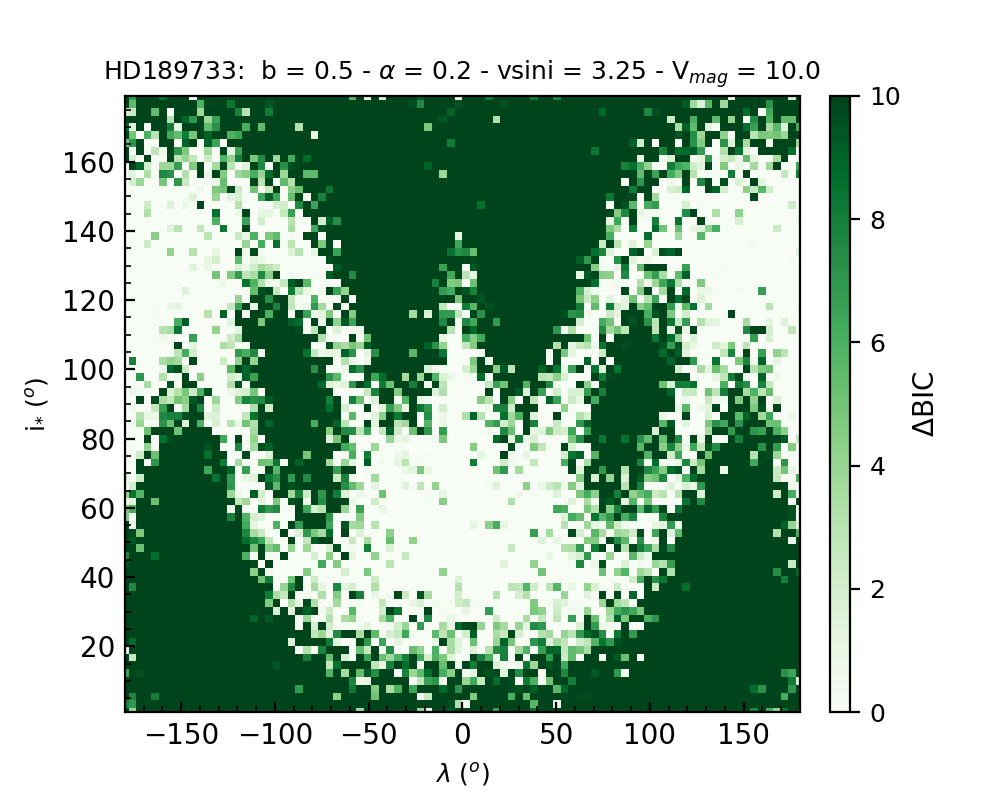}
        \caption{Two-dimensional heat maps illustrating the preferred parameter space (colour-coded by the BIC) for  detecting stellar DR as a function of projected obliquity and stellar inclination for two different sampling rates (left: 100s, right: 300s). These results have been obtained with $b=0.5$, $V=10$ and $\alpha=0.2$, with the other parameters set to the values listed in Table~\ref{tab:param}.}
        \label{hd189nocb}
    \end{figure*}

    We assumed an exposure time of 100 seconds for our simulated data so far, which is a fairly optimistic assumption. Figure~\ref{hd189nocb} shows the expected difference between this 100~s cadence and that of a 300~s exposure time, which is more in line with expectations for empirical observations of one transit for a bright target.

    Increasing the exposure time increases the S/N and should yield better results, but the decreased sampling rate balances the increase in accuracy. Even an exposure time of 600s only leads to marginally better results. Furthermore, a long exposure time runs the risk of not sampling ingress and egress, which is more detrimental to DR detection than the high S/N. Integrating readout delay in the simulation has the effect of decreasing the sampling rate without increasing exposure time, but does not produce any significant difference.

    \subsubsection{Impact of brightness}
    \label{subsubsec:brightness_nocb}

    Our ability to prefer a DR model is driven by both the precision and sampling rate of our measurements with respect to the absolute magnitude of the DR. Increases in the S/N will naturally lead to an increase in the measured precision on the local velocities. Hence, we explored the potential impact of the $V$ magnitude by scaling our uncertainties following Equation~\ref{eqn:scale}. In particular, we investigated scenarios for $V=10$ and $V=12$, which values are still considered fairly bright for transiting exoplanet systems, but are substantially dimmer than an HD~189733-like system ($V\approx7.7$).

    \begin{figure}[h]
    \centering
    \includegraphics[trim=3.5cm 1cm 1.4cm 1.5cm, clip, scale=0.43]{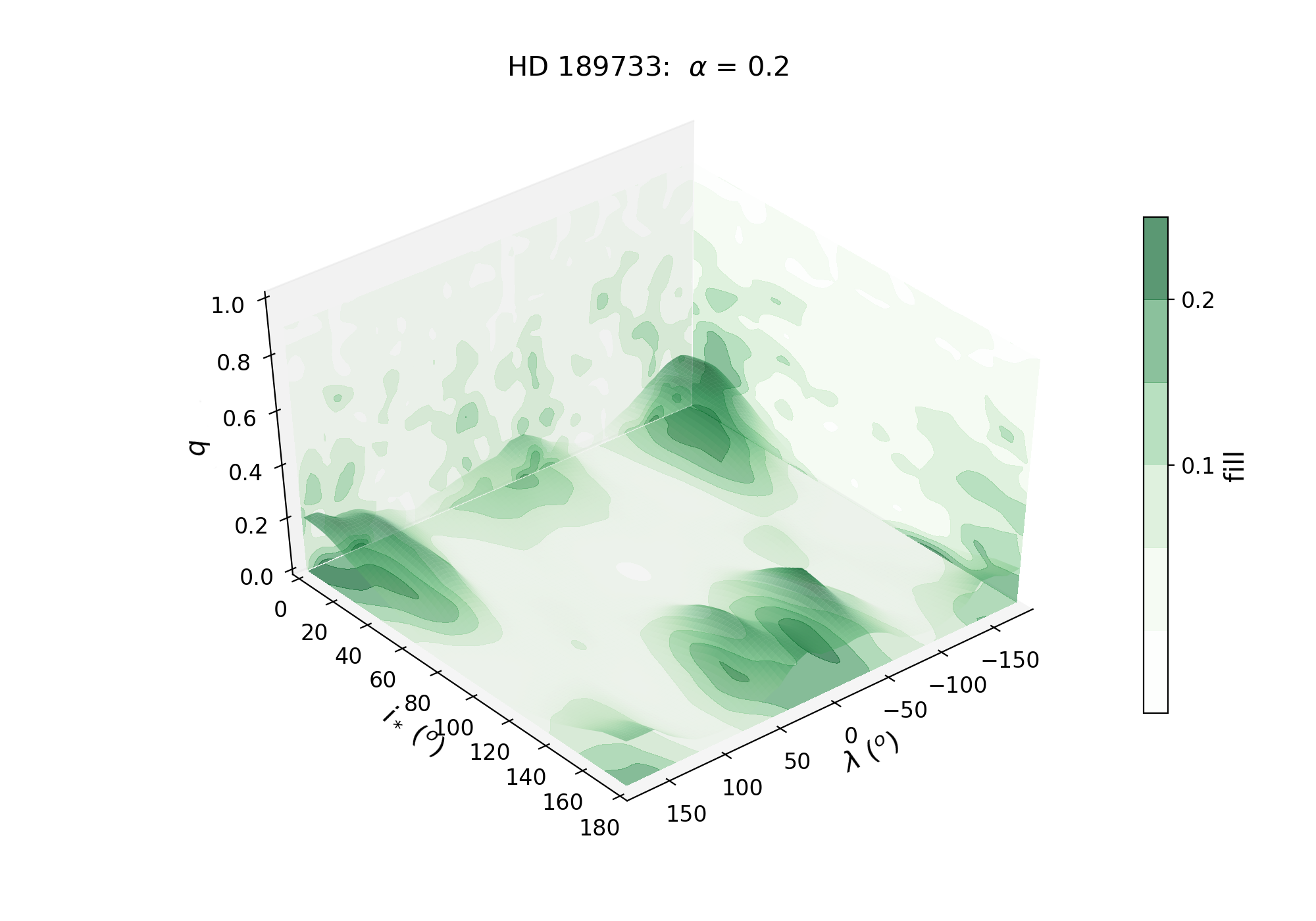}
        \caption{Three-dimensional heat map of the likelihood of detecting DR, using HD~189733 system parameters and $V = 10$, with a HARPS-like precision. These results have been obtained with $\alpha=0.2$. The parameters that were held constant are listed in Table~\ref{tab:param}.}
        \label{hd189733fullvmag10H}
    \end{figure}

    \begin{figure}[h]
    \centering
    \includegraphics[trim=3.5cm 1cm 1.4cm 1.5cm, clip, scale=0.43]{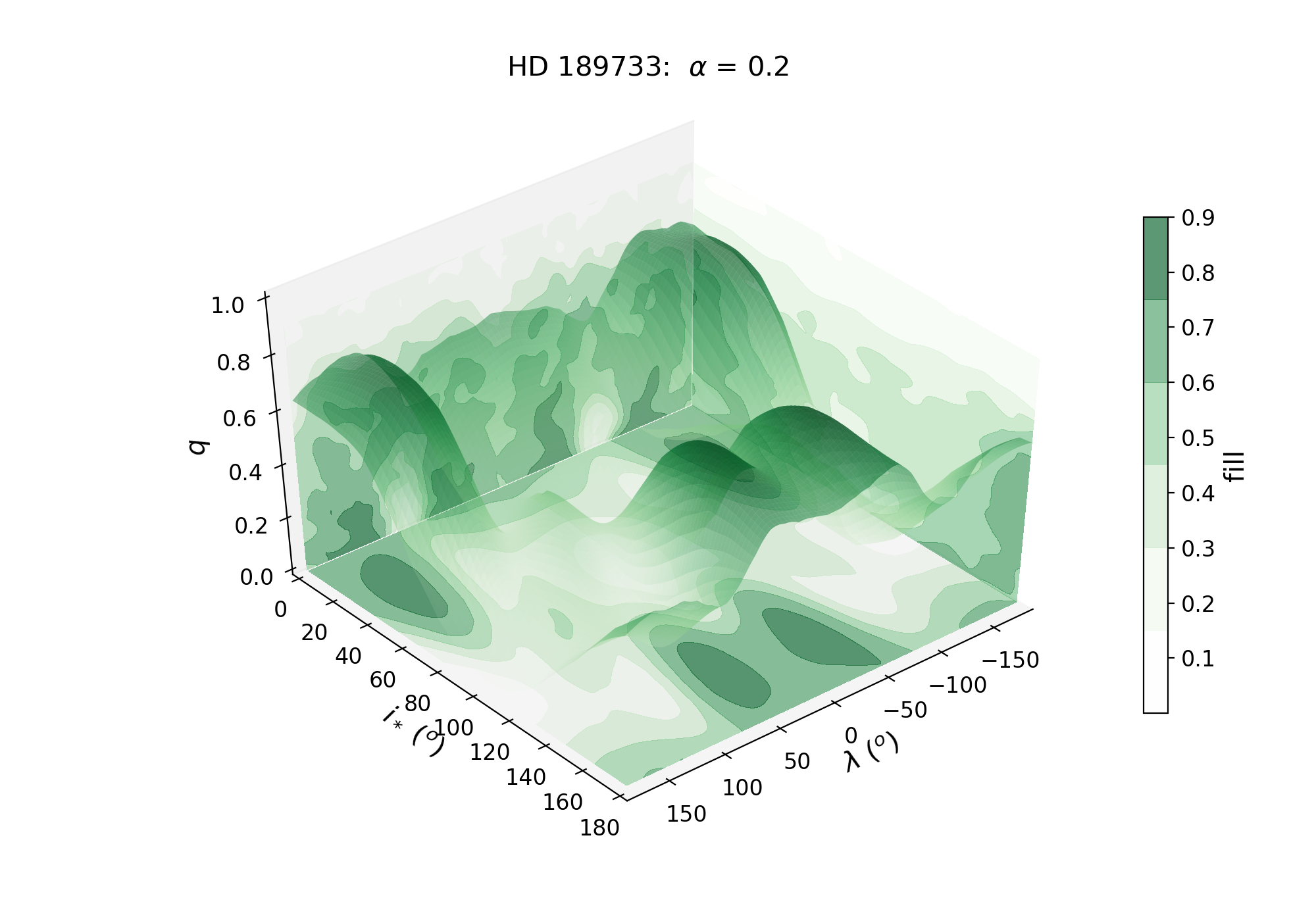}
        \caption{Three-dimensional heat map of the likelihood of detecting DR, using HD 189733 system parameters and $V = 10$, with an ESPRESSO-like precision. These results have been obtained with $\alpha=0.2$. The parameters that were held constant are listed in Table~\ref{tab:param}.}
        \label{hd189733fullvmag10E}
    \end{figure}

\begin{figure}[h]
    \centering
    \includegraphics[trim=3.5cm 1cm 1.4cm 1.5cm, clip, scale=0.43]{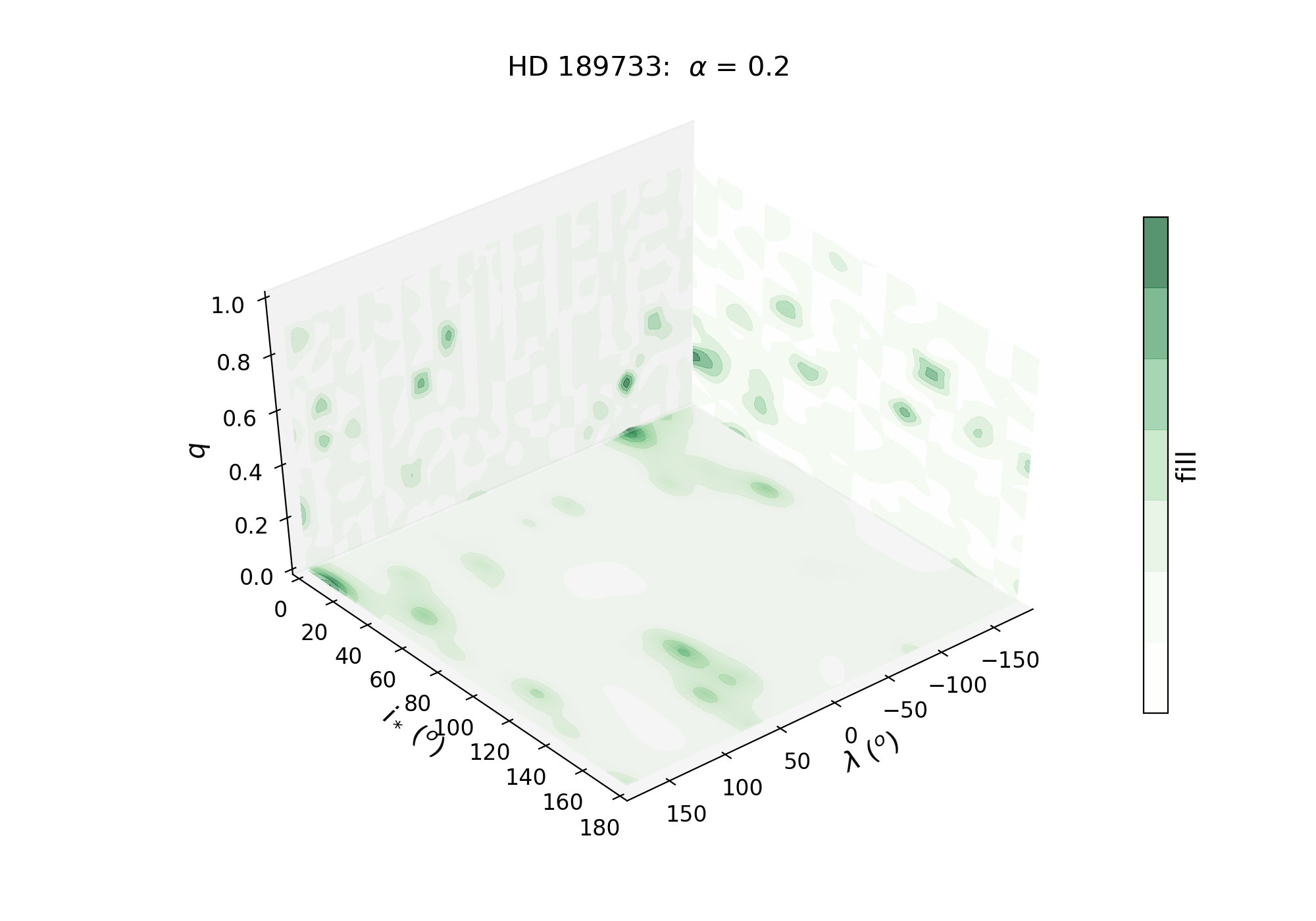}
        \caption{Three-dimensional heat map of the likelihood of detecting DR, using HD 189733 system parameters and $V = 12$, with an ESPRESSO-like precision. These results have been obtained with $\alpha=0.2$. The parameters that were held constant are listed in Table~\ref{tab:param}.}
        \label{hd189733fullvmag12E}
    \end{figure}

    Figure~\ref{hd189733fullvmag10H} demonstrates that it may be extremely difficult to measure a solar-like DR on even $V = 10$ systems using a HARPS-like setup, at least given the cadence herein. Detecting DR in these conditions would be exceedingly hard for any impact parameter $b > 0.2$. However, if we assume an ESPRESSO/VLT setup, we obtain very similar results as before (i.e. a V=7.7 host star observed with a HARPS-like setup, see Figure~\ref{hd189733fullvmag10E}); this illustrates the powerful impact of increasing the precision through a collection of resolving power and increased instrumental precision. Figure~\ref{hd189733fullvmag10E} is remarkably analogous to Figure~\ref{hd189733fullv3.5}. Decreasing the magnitude further and setting $V = 12$, we now approach the limit of an ESPRESSO/VLT-like setup, as shown in Figure~\ref{hd189733fullvmag12E}. This last figure is almost identical to Figure~\ref{hd189733fullvmag10H}, in that detecting DR in such a system seems to be beyond reach for now, at least with one transit.

    \subsubsection{Impact of the stellar rotation rate}
    \label{subsubsec:rotation_rate_nocb}

\begin{figure}[h]
    \centering
    \includegraphics[trim=3.5cm 1cm 1.4cm 1.5cm, clip, scale=0.43]{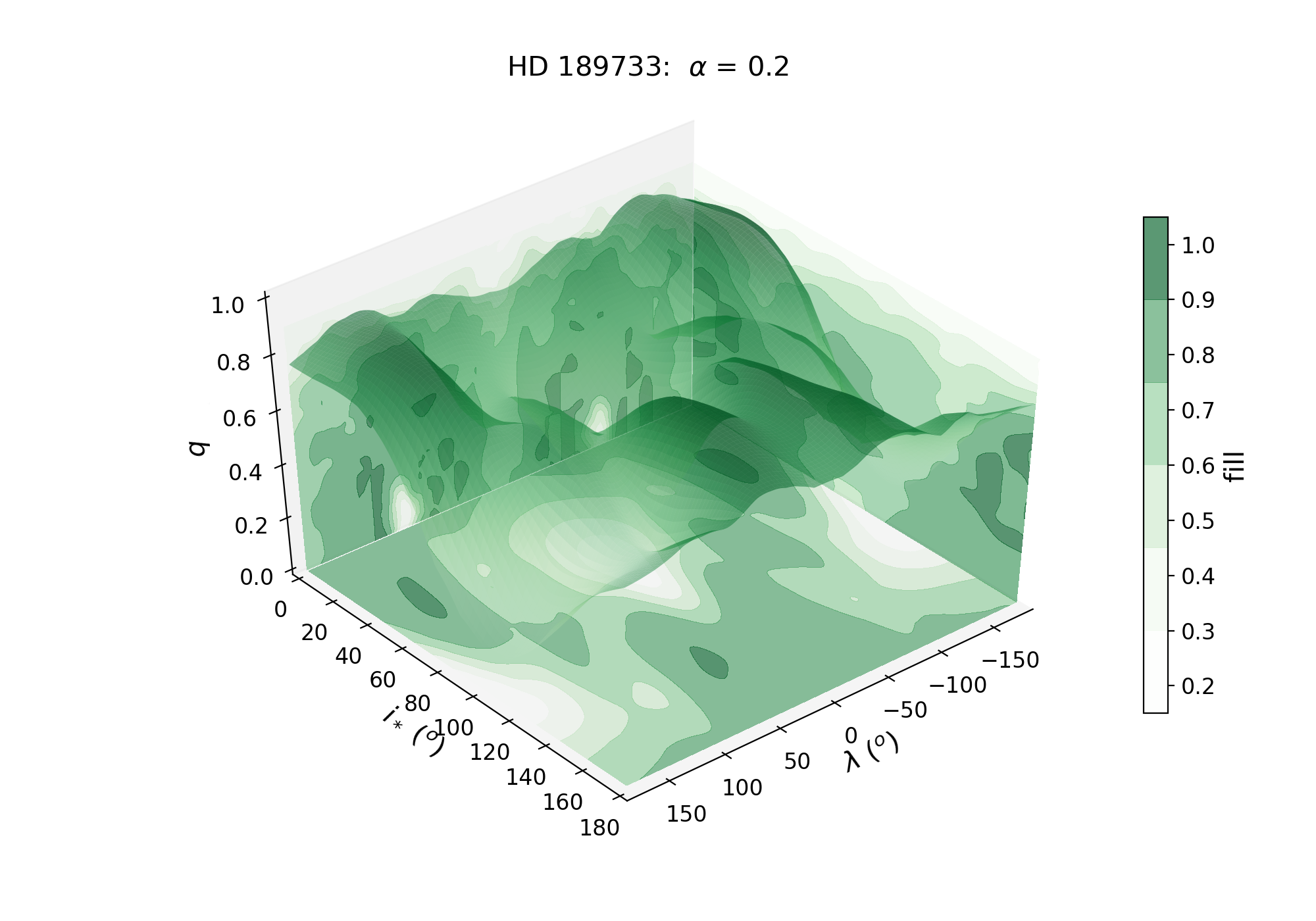}
        \caption{Three-dimensional heat map of the likelihood of detecting DR, using HD 189733 system parameters, $v_{eq}\sin i_* = 15$~km~s$^{-1}$ and $V = 10$, with a HARPS-like precision. These results have been obtained with $\alpha=0.2$. The parameters that were held constant are listed in Table~\ref{tab:param}.}
        \label{hd189733fullvmag10H2}
    \end{figure}

    Alternatively, it is also easier to detect DR when the absolute magnitude of the effect is larger. Hence the limits on brightness and instrumental setup or precision can be somewhat mitigated depending on the $v_{eq}\sin i_*$ of the target. Figures~\ref{hd189733fullvmag10H} - \ref{hd189733fullvmag12E} have all been obtained using the $v_{eq}\sin i_*$ listed in Table~\ref{tab:param}, and we demonstrate that while an ESPRESSO/VLT like setup is viable for slowly rotating, dim stars (up until $V\approx12$), the same cannot be said of HARPS. However, if we consider a much faster rotator (i.e. $v_{eq}\sin i_*=15~$km~s$^{-1}$), a HARPS-like setup becomes feasible, as Figure~\ref{hd189733fullvmag10H2} shows.

    It is interesting to see in Figure~\ref{hd189733fullvmag10H2} that while the crests reach almost as high as they did previously, in the nominal HD~189733 case (cf. Figure~\ref{hd189733fullv3.5}), the troughs in the 3D surface are much deeper. This suggests that while DR detection is definitely possible in the case of a dim but fast rotator using a HARPS-like setup, it is much more sensitive to the parameters of the star-planet orientation, and our ability to retrieve DR may be limited to a much reduced set of targets.

\subsection{Stellar rotation and centre-to-limb convection variation}
\label{Sect:resultscb}
    In this second part, we now consider the compounded impact of both DR and CLV. We are interested in determining whether DR+CLV or SB+CLV is preferred and for which parameter space. We also examine whether SB+CLV can be confused for pure DR and pure DR be confused for pure CLV.

\begin{figure*}[h]
    \centering
    \includegraphics[trim=0mm 0mm 0mm 11.1mm,clip,width=6cm]{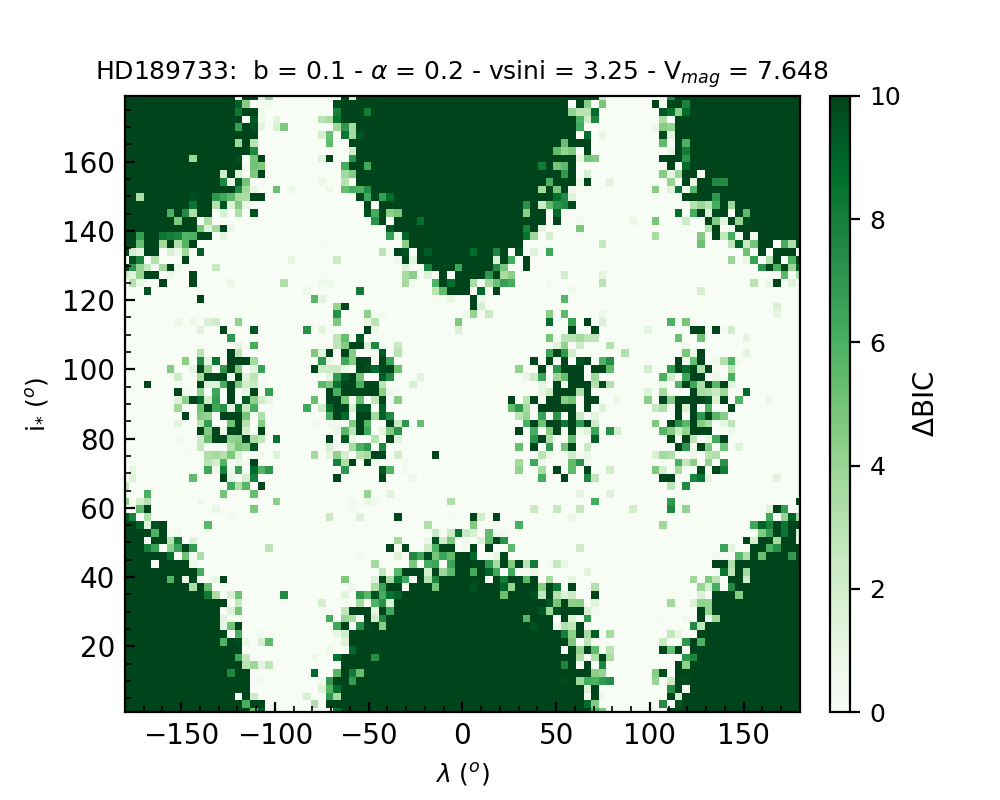}
    \includegraphics[trim=0mm 0mm 0mm 11.1mm,clip,width=6cm]{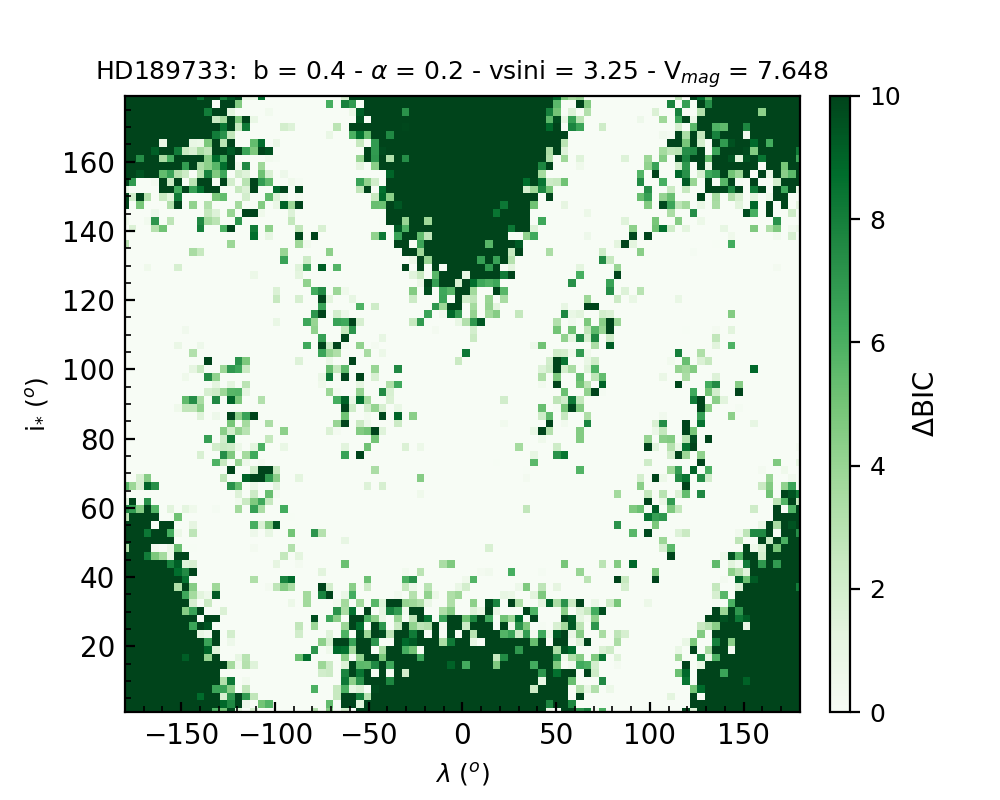}
    \includegraphics[trim=0mm 0mm 0mm 11.1mm,clip,width=6cm]{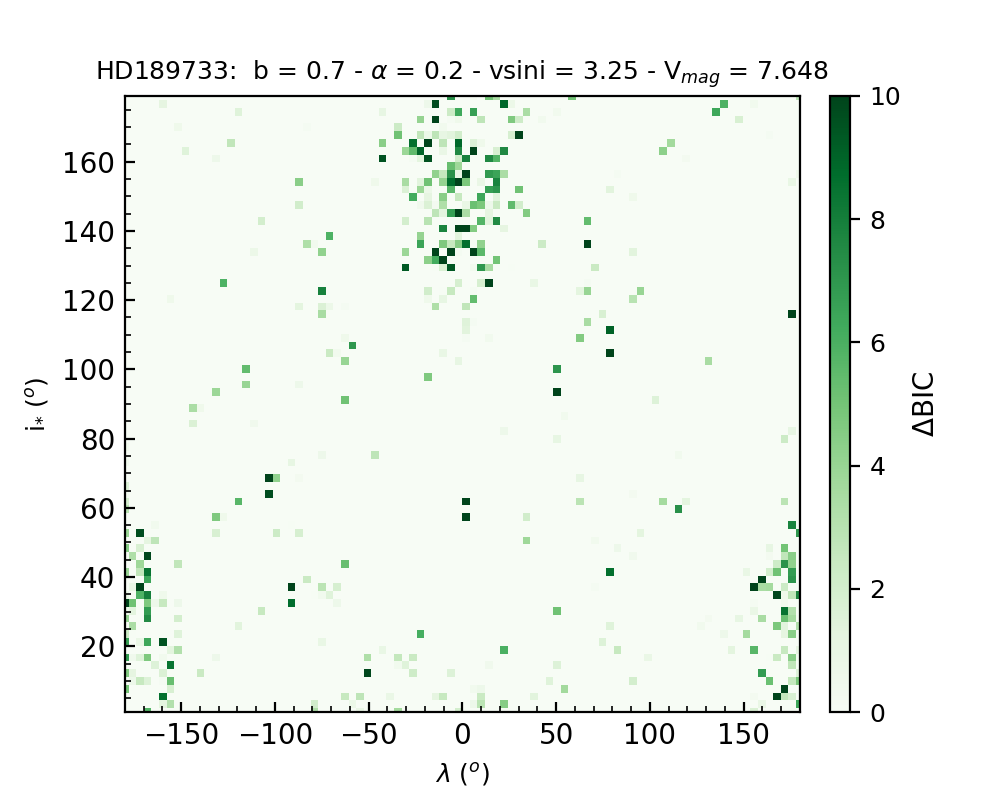}
        \caption{Two-dimensional heat maps of the degree of preference for the DR+CLV model over the SB+CLV model, depending on the impact parameter $b$ ($b = 0.1, 0.4, \text{and }0.7$ from left to right). These results have been obtained with $\alpha=0.2$, and the parameters that were held constant across the three figures are listed in Table~\ref{tab:param}.}
        \label{hd189cbnoalpha}
    \end{figure*}

    Starting with the first question, we evaluated the likelihood of retrieving the injected data (DR+CLV) as opposed to the simpler model (SB+CLV). In this configuration, we recall that the DR+CLV model has six free parameters ($\lambda$, $i_*$, $v_{eq}$ , and $\alpha$  from the DR model described in Section~\ref{subsec:Rot_Only}, plus $c_1$ and $c_2$), while the SB+CLV model has four ($\lambda$ and $v_{eq}\sin i_*$, plus $c_1$ and $c_2$ as well). In the same way as for the case without CLV (cf. Section~\ref{subsec:Rot_Only}), Figure~\ref{hd189cbnoalpha} illustrates the preference for DR+CLV over SB+CLV for three values of the impact parameter $b$.

    As before, the darker the shade of green, the more likely it is that we can retrieve the injected data. The main hot-spots to retrieve DR+CLV are

    \begin{itemize}
        \item $i_*>110^{\rm{o}}$ and $|\lambda|<50^{\rm{o}}$
        \item $i_*<50^{\rm{o}}$ and $|\lambda|>100^{\rm{o}}$
        \item $i_*\approx90^{\rm{o}}$ and $|\lambda|\approx60^{\rm{o}}$
        \item $i_*\approx90^{\rm{o}}$ and $|\lambda|\approx120^{\rm{o}}$.
    \end{itemize}

    Despite similarities with the results from the previous section, we observe some important changes around $i_*\approx90^{\rm{o}}$ and $\lambda\approx\pm90^{\rm{o}}$: Whereas we observed two distinct hot-spots in Figure~\ref{hd189nocbnoalpha} (without CLV) that appeared at moderate to high impact parameters, now  a more complex structure emerges. Instead of the two ellipses around $\lambda\approx\pm90^{\rm{o}}$ and $60^{\rm{o}}<i_*<120^{\rm{o}}$,  four arms appear to wrap around the same parameter space, and the centre area ($i_*=90^{\rm{o}}$ and $\lambda=\pm90^{\rm{o}}$) remains empty. This result might be a sign that at this point, DR is being confused with the CLV and that consequently, the less complicated SB+CLV model is the preferred fit to the data points for this specific orientation.

    Another major difference is that without CLV, these same hot-spots around $i_*\approx90^{\rm{o}}$ and $\lambda\approx\pm90^{\rm{o}}$ can still be observed at high impact parameters (cf. right plot of Figure~\ref{hd189nocbnoalpha}). However, when CLV is considered, the right plot of Figure~\ref{hd189cbnoalpha} shows that the four arms fade away at a lower impact parameter compared to the other hot-spots that are present in the centre plot of the same figure. Hence, when CLV is present in the injected data, retrieving it along with DR is more difficult the higher the impact parameter.

\begin{figure}[h]
    \centering
    \includegraphics[trim=3.5cm 1cm 1.4cm 1.5cm, clip, scale=0.43]{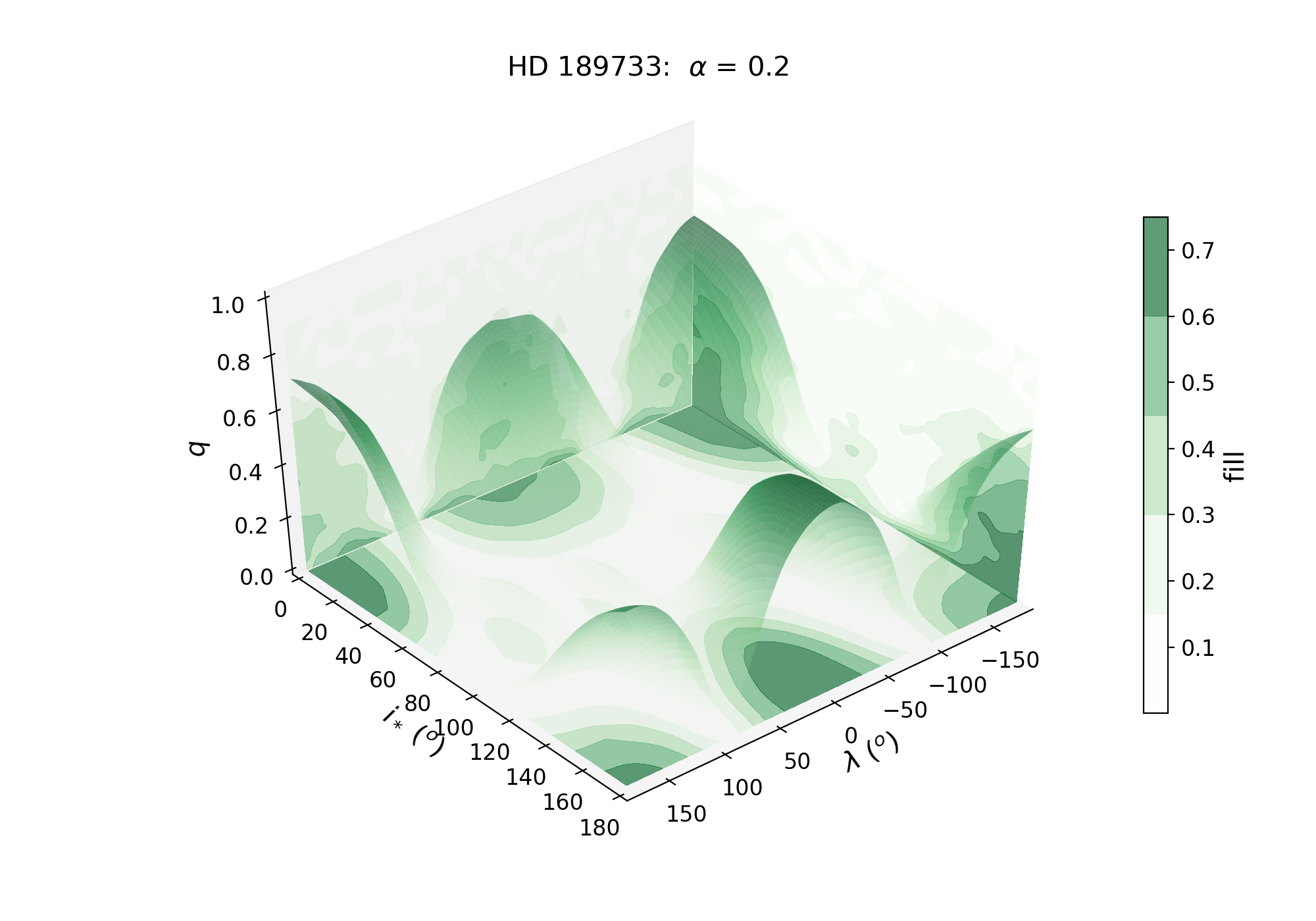}
        \caption{Three-dimensional heat map of the likelihood of retrieving DR+CLV as compared to SB+CLV. A darker colour indicates a lower preference for DR+CLV in the projected area. These results have been obtained with $\alpha=0.2$, and the parameters that were held constant are listed in Table~\ref{tab:param}.}
        \label{hd189cbfull}
    \end{figure}

    The 3D heat map (cf. Figure~\ref{hd189cbfull}) shows that the density map in the bottom panel is more contrasted than the case without CLV (cf. Figure~\ref{hd189733fullv3.5}). While DR+CLV can be reliably detected up to high values of $b$ around particular hot-spots, it is much less favoured between them, even at low impact parameters: there are dead spots in the parameter space when CLV is considered that were not present without CLV. This means that our ability to detect DR is much more dependent on the orientation of the system than before.

    \subsubsection{Accuracy of fits}
    \label{subsubsec:accuracy_CLV}

    \begin{figure*}[h]
    \centering
    \includegraphics[trim=0mm 0mm 0mm 11.1mm,clip,width=6cm]{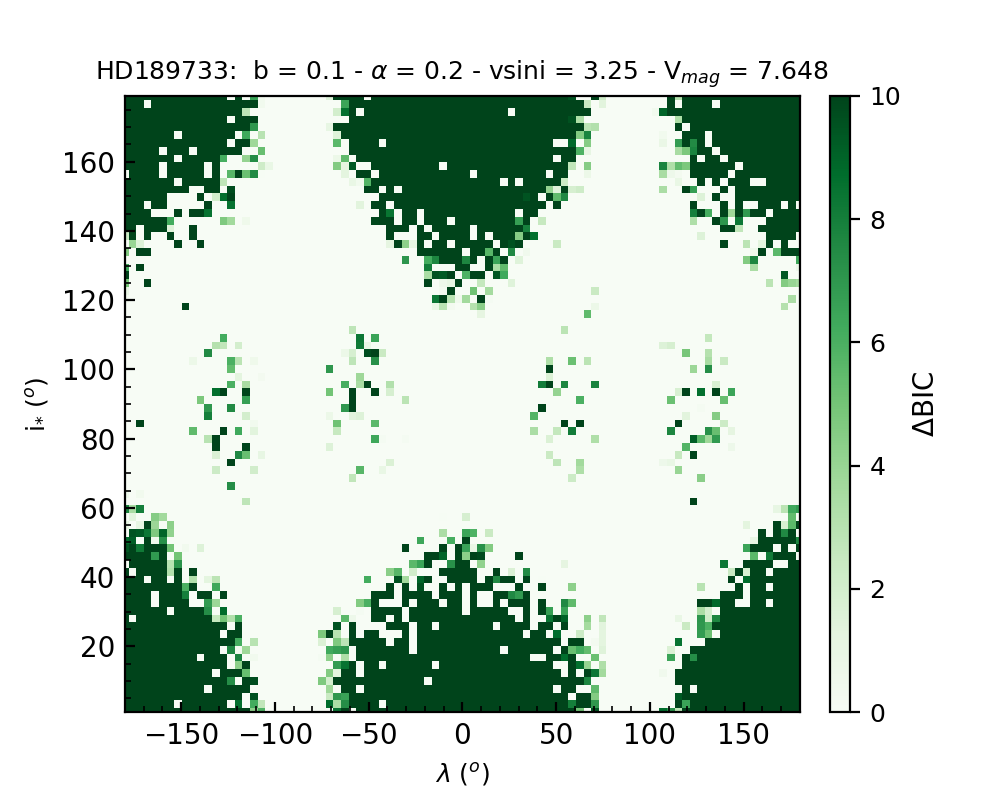}
    \includegraphics[trim=0mm 0mm 0mm 11.1mm,clip,width=6cm]{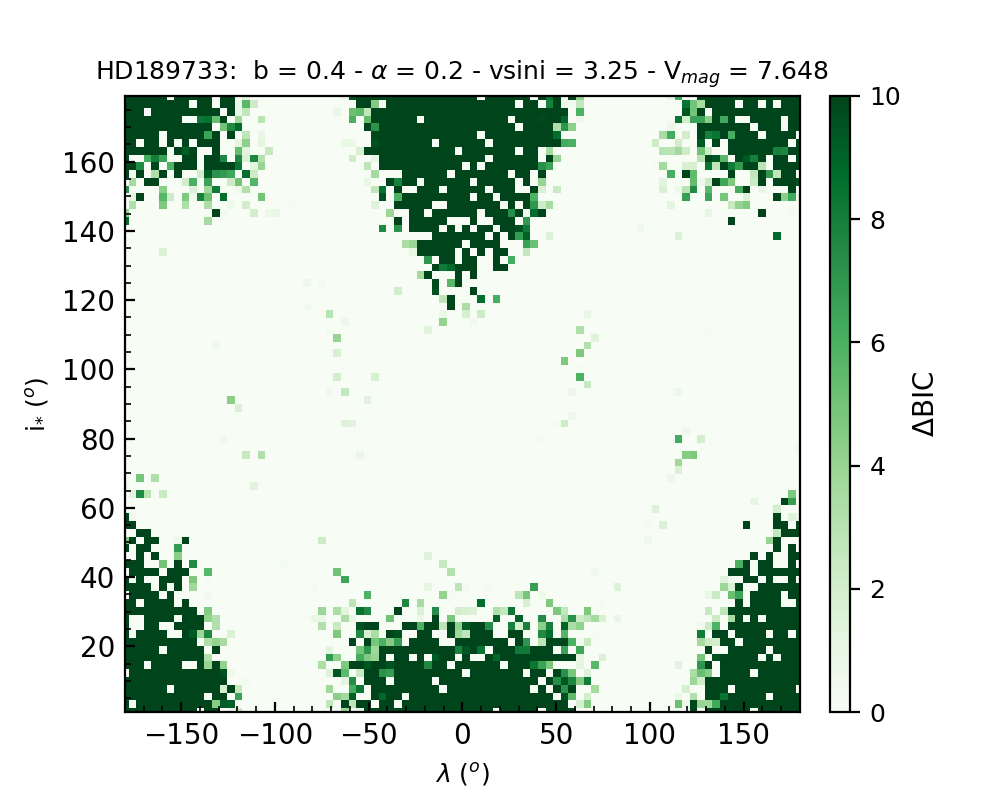}
    \includegraphics[trim=0mm 0mm 0mm 11.1mm,clip,width=6cm]{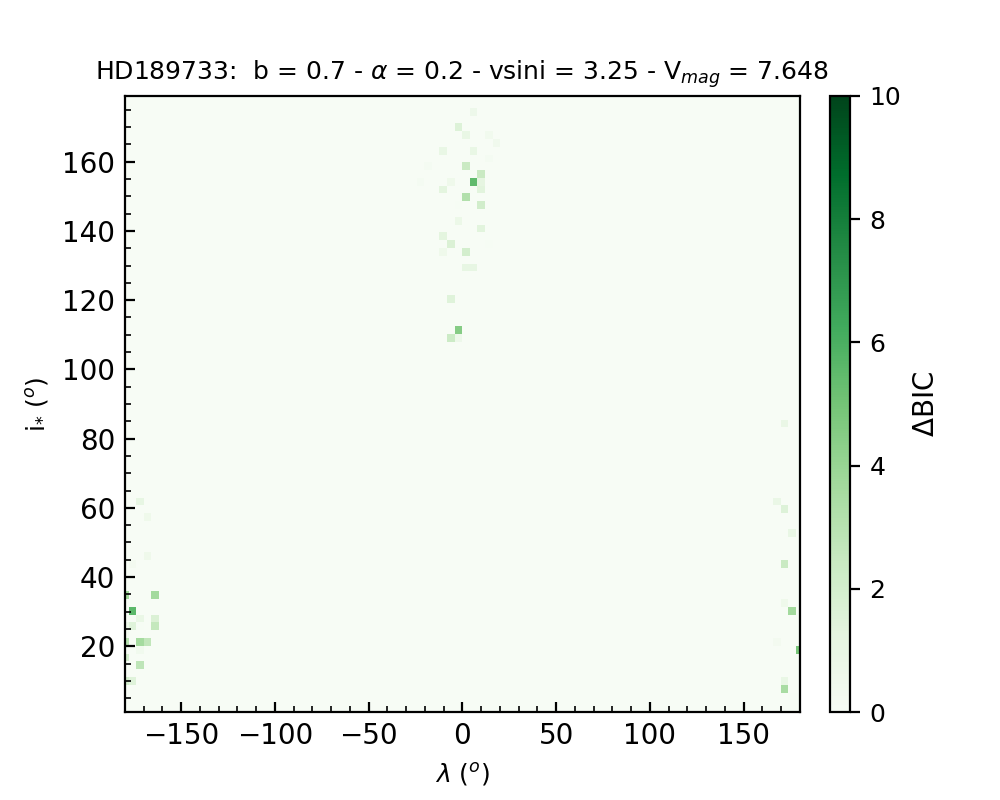}
        \caption{DR+CLV preference over SB+CLV depending on $b$ ($b = 0.1, 0.4, \text{and }0.7$ from left to right), considering the proximity of the fitted value of $\alpha$ to the injected value. Accurate detections are those where the recovered $\alpha$ agrees with the injected value within 0.05. These results have been obtained with $\alpha=0.2$, and the parameters that were held constant in the three figures are listed in Table~\ref{tab:param}.}
        \label{hd189cbalpha}
    \end{figure*}

    Like the previous subsection (cf Section~\ref{subsec:Rot_Only}), we are interested in the accuracy of the fit on $\alpha$ where DR+CLV is preferred to SB+CLV. We compare Figure~\ref{hd189cbnoalpha} and Figure~\ref{hd189cbalpha} to determine whether the hot-spots that we previously identified result in a reliable detection of the differential rotation. The area in which both the DR+CLV model is preferred and the difference between the fitted value of $\alpha$ and the injected value is less than $0.05$, is less expansive than in the first series of plots (DR-only). The sharpness of the aforementioned hot-spots declines sharply: while the underlying structure is very similar between Figure~\ref{hd189cbnoalpha} and Figure~\ref{hd189cbalpha}, it becomes more fuzzy in the latter. The conclusion that we drew in the DR-only case seems to apply in this case as well: a particular area might be favourable to retrieve the injected model on average, but given a single point in the parameter space, the fit might not be reliable.

    To verify whether the CLV could be mistaken for DR and introduce a bias in the fit on $\alpha$, we ran several sets of simulations (see Table~\ref{tab:param} for the system configuration), injecting SB+CLV (setting the parameter $\alpha=0$). In the first case, we compared a SB fit with a DR fit. Our results show that without considering the accuracy of the fit on the value of $\alpha$, a DR model explains the injected SB+CLV better than the SB model and could therefore point towards a true SB+CLV being confused for DR (if the CLV contribution to the fit were not considered). We suggest that this ambiguity would certainly be lifted if the fitted values of $\alpha$ could be checked against a credible estimate of the true value. However, the challenge is indeed that a sensible estimation of the true value of $\alpha$ is needed to begin with, which is a problematic prospect when constraining the value of this parameter is the primary intention of the study.

    Interestingly, despite the inconclusive results of this approach, almost all of the best fits we obtained when trying to fit an injected SB+CLV with a DR model led to a recovered $\alpha=\pm1$. Such extreme results should certainly be interpreted with caution and might point towards the conclusion that trying to fit a pure DR model on SB+CLV will produce inadequate effects.
    In the second scenario, we take our case with the injected SB+CLV velocities and compare an SB+CLV model fit with a DR model (i.e. SB+CLV vs DR, as opposed to the previous SB vs DR test), which is interesting because both models have the same number of free parameters. Without imposing any conditions on any parameters, we find that the DR fit cannot explain the data better than the SB+CLV fit. Hence, if the true stellar surface is explained by SB+CLV, it may be possible to prefer a DR model over an SB model (albeit an inaccurate DR), but the best fit will still be from a SB+CLV model. Regardless of fitting an SB, DR, or SB+CLV model to our model star (with SB+CLV injected), the recovered projected obliquity remained robust.
    From these results, we conclude that while the possibility of CLV being mistaken for DR cannot completely be ruled out, it seems to be very limited. For the accuracy of the fits on $\lambda$, $i_*$ , and $v_{eq}$, our results are in line with those we obtained previously. We are able to retrieve the injected parameters with good precision: The mean offset between the fitted and the injected value for $\lambda$ is slightly worse than in Section~\ref{subsubsec:accuracy_DR} at $-0.48^{\rm{o}}$ with a standard deviation of $13.45^{\rm{o}}$. The increased width of the offset distribution shows that the fit is not quite as reliable as in the DR case. For $i_*$, we find a mean offset of $0.16^{\rm{o}}$ with a standard deviation of $20.02^{\rm{o}}$. Despite a smaller mean offset on $v_{eq}$ at $-1.5$ km~s$^{-1}$ (standard deviation $26.51$ km~s$^{-1}$) compared to Section~\ref{subsubsec:accuracy_DR}, the mean reconstructed value for $v_{eq}{\sin i_{*}}$ is farther away from the injected value, at $3.57$ km~s$^{-1}$ (standard deviation $3.12$ km~s$^{-1}$) instead of $3.25$ km~s$^{-1}$ (cf. Table~\ref{tab:param}). Lastly, the fitted $\alpha$, despite a mean offset to the injected value of $-0.05$ (standard deviation $0.18$), is not as reliable as in the DR-only case, especially at high impact parameters (cf. Figure~\ref{hd189cbalpha}).

   \subsubsection{Impact of the sampling rate}

\begin{figure*}[h]
    \centering
    \includegraphics[trim=0mm 0mm 0mm 11.1mm,clip,width=8.5cm]{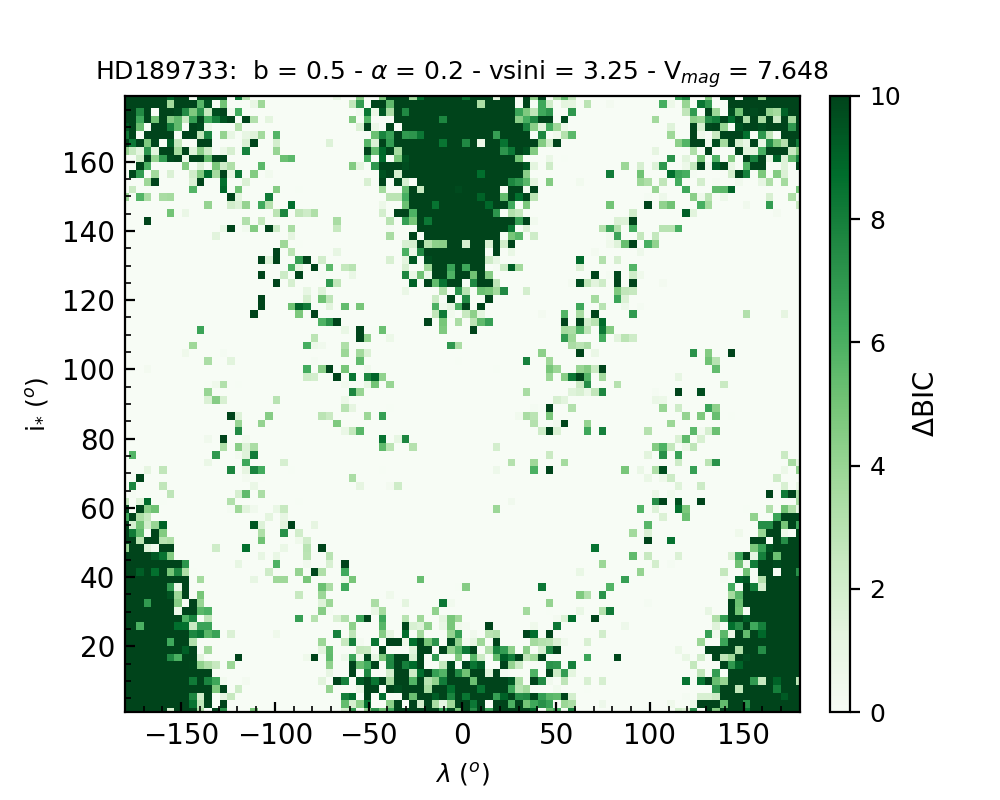}
    \includegraphics[trim=0mm 0mm 0mm 11.1mm,clip,width=8.5cm]{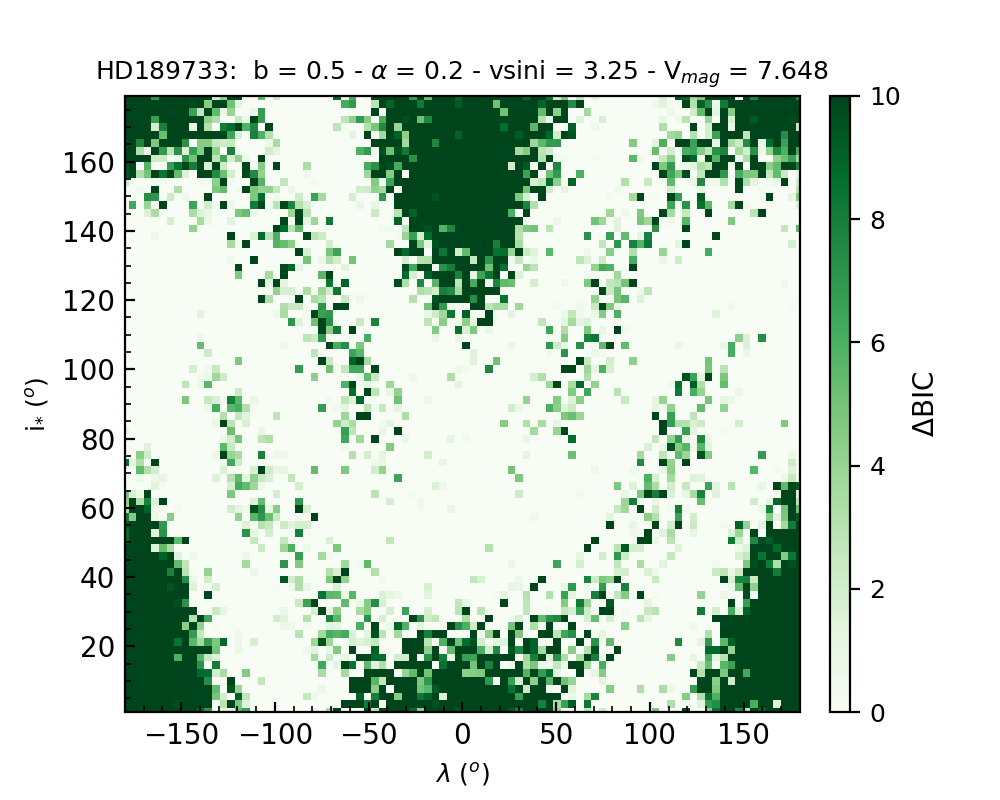}
        \caption{Two-dimensional heat maps of the parameter space ($b=0.5$), with the preference for DR+CLV over SB+CLV colour-coded in green for two different sampling rates (left: 100s, right: 300s). These results have been obtained with $\alpha=0.2$, and the parameters that were held constant across the three figures are listed in Table~\ref{tab:param}.}
        \label{hd189cb}
    \end{figure*}

    When we examined the impact of exposure time or cadence on our ability to retrieve DR+CLV, our observations are largely the same as in the DR only case. We conclude that while the increase in S/N resulting from the longer exposure time helps, the benefits are largely negated by the lower sampling rate. Figure~\ref{hd189cb} illustrates this point.

    Complementary results support the fact that while the two effects we discussed cancel out on a single transit, obtaining multiple transits helps to avoid the negative effect of a lower sampling rate while maintaining the full benefit of the higher S/N.

    \subsubsection{Impact of brightness and stellar rotation rate}

    For the influence of brightness and the rotation rate for the preference for the DR+CLV over the SB+CLV fit, our conclusions are once again very similar to the first section. A drop in brightness to $V=10$ or $V=12$ would prove very challenging, even considering the precision of an ESPRESSO/VLT setup.

    \begin{figure}[h]
    \centering
    \includegraphics[trim=3.5cm 1cm 1.4cm 1.5cm, clip, scale=0.43]{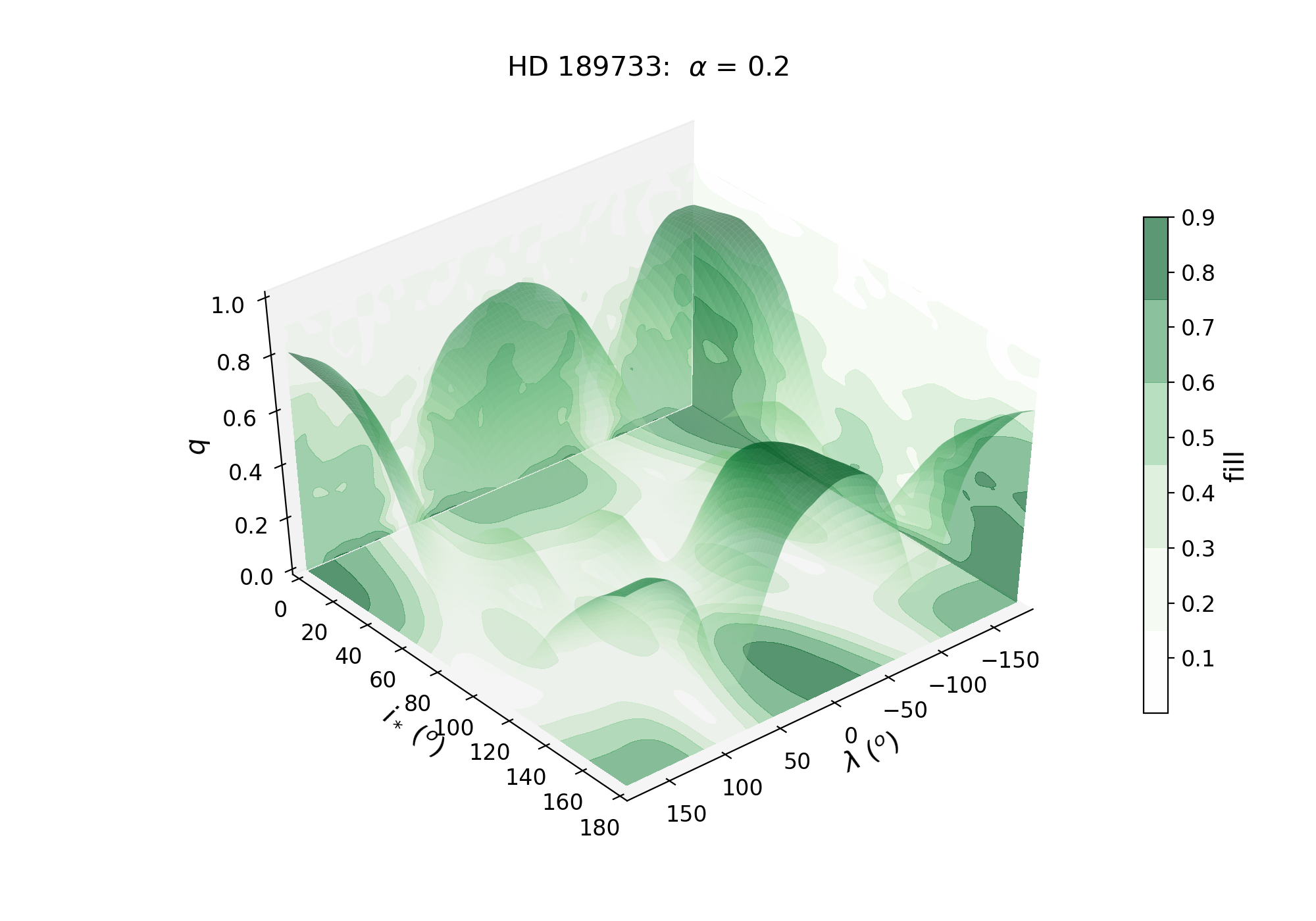}
        \caption{Three-dimensional heat map of the likelihood of detecting DR+CLV, using HD 189733 system parameters, $v_{eq}\sin i_* = 15$~km~s$^{-1}$ and $V = 10$, with a HARPS-like precision. These results have been obtained with $\alpha=0.2$, and the parameters that were held constant are listed in Table~\ref{tab:param}.}
        \label{hd189733cbfullvmag10H2}
    \end{figure}

    Comparison of Figure~\ref{hd189733cbfullvmag10H2} with Figure~\ref{hd189733fullvmag10H2} shows that if CLV is present in the injected data (at the rate injected here, following a solar MHD simulation of the Fe I 6302~\r{A}\footnote{We recall that hotter stars have a higher expected CLV, but also higher temporal variations from granulation or p-modes.}), a solar-like DR is unlikely to be retrieved in a practical way using a HARPS-like setup if the star is not observed pole-on or pole-away, even in the case of a fast rotator ($v_{eq}\sin i_*=15~$km~s$^{-1}$ in both figures). The fraction of the parameter space in which the system orientation favours detection is much more limited when CLV is injected in the model.


\section{Conclusion}
\label{Sect:conc}
    To take full advantage of the improved precision of recent instruments (e.g. ESPRESSO), a more detailed understanding of the stellar surface is important. Throughout this paper, we have presented our results for the optimal parameter space for detecting stellar differential rotation and centre-to-limb convective variations. To do this, we simulated a typical transiting hot Jupiter (similar to HD~189733) and varied several parameters, including the projected obliquity, stellar inclination, and impact parameter.

    In the first part of this paper, we examined the likelihood of retrieving an accurate estimate of the differential rotational sheer $\alpha$ depending on system parameters, leaving out any contribution of the CLV in the injected data. We find that DR detection is null for $\lambda=\pm90^{\rm{o}}$ at low impact parameters ($\sim b\in[0,0.2]$) for all $i_*$. The main hot-spots for detection are $120^{\rm{o}}<|\lambda|<180^{\rm{o}}$ for $i_*<90^{\rm{o}}$, and $|\lambda|<60^{\rm{o}}$ for $i_*>90^{\rm{o}}$ on average, with the planet transiting the lower half of the star in our coordinate system. Additionally, at moderate impact parameters, DR detection is likely at $|\lambda|\approx90^{\rm{o}}$ for $i_*\approx90^{\rm{o}}$, which corresponds to the configuration in which the largest number of latitudes are crossed. At higher impact parameters ($\sim b>0.7$), DR detection is increasingly difficult even in the regions described above. We also find that the estimate on $\alpha$ is reliable and only drops in accuracy near the edges of the hot-spots and with increasing values of $b$.

    We also observe that increasing the exposure time does not drastically improve detection, as the increase in S/N is offset by the loss of sampling. Unsurprisingly, multiple transits will alleviate the adverse impact of the lower sampling rate.

    Regarding the influence of brightness and rotation rate, we observe that the detection limit for a HARPS-like setup is probably about $V=10$ for a moderately slow rotator such as HD~189733, while $V=12$ seems to be the limit considering an ESPRESSO-like setup. If we consider faster rotators (e.g. $\sim$15~km~s$^{-1}$), a HARPS-like setup may detect DR beyond $V=10$ (if it is photon-noise dominated), although within a more limited parameter space.

    In the second part of this paper, we investigated the impact of a CLV component on the injected data. Similarly to the first case, detection still is compromised for $\lambda=\pm90^{\rm{o}}$ at low impact parameters (for all $i_*$). However, the emerging structure in the parameter space is different and more complex than previously. The main hot-spots for detecting DR+CLV are $|\lambda|>100^{\rm{o}}$ for $i_*<50^{\rm{o}}$ and $|\lambda|<50^{\rm{o}}$ for $i_*>110^{\rm{o}}$, which are visible up until $b=0.7$, that is slightly worse than in the first case. Additionally, we observe secondary hot-spots at $|\lambda|<50^{\rm{o}}$ for $i_*<60^{\rm{o}}$, $|\lambda|>110^{\rm{o}}$ for $i_*>120^{\rm{o}}$, as well as $|\lambda|\approx60^{\rm{o}}$ and $|\lambda|\approx120^{\rm{o}}$ for $i_*\approx90^{\rm{o}}$, which subside at lower impact parameter than the others as $b$ increases. Surprisingly, DR+CLV detection is null in the region around $|\lambda|\approx90^{\rm{o}}$ and $i_*\approx90^{\rm{o}}$ that we identified in the first part, maybe due to degeneracy between the two components.

    When comparing the fitted value on $\alpha$ with the injected parameter, we find that the hot-spots near $i_*=90^{\rm{o}}$ are practically non-existent at any impact parameter, constraining DR+CLV detection to the edges of the parameter space. The other conclusions we reached in the first part regarding the influence of brightness, exposure time, and rotation rate are still applicable, with the major difference being that reliable detection of DR+CLV is confined to more well-delimited regions of the parameter space, and to lower impact parameters overall. For these reasons, detection for $V>10$ might prove challenging even with an ESPRESSO-like setup, especially if the number of available transits is limited.

    Lastly, we established that CLV is unlikely to be mistaken for DR if reasonable constraints can be placed on the value of $\alpha$. It seems that when CLV and DR might be confused, the DR fit produces largely unreliable results that would not pass scrutiny.

\begin{acknowledgements}
      The authors acknowledge the financial support of the National Centre for Competence in Research PlanetS supported by the Swiss National Science Foundation (SNSF). This research has made use of NASA’s Astrophysics Data System Bibliographic Services. HMC was supported during this project from both an NCCR PlanetS CHEOPS Fellowship and a UKRI Future Leaders Fellowship (MR/S035214/1). VB acknowledges funding from the European Research Council (ERC) under the European Union's Horizon 2020 research and innovation programme (project {\sc Spice Dune}, grant agreement No 947634).
\end{acknowledgements}


\bibliographystyle{aa}
\bibliography{mybib}

\end{document}